\documentclass{article}
\usepackage{indentfirst}
\usepackage[utf8]{inputenc}
\usepackage{cite}
\usepackage{array}
\usepackage{float} 
\usepackage[table,xcdraw]{xcolor} 
\usepackage{booktabs} 
\usepackage{caption} 
\usepackage{amsmath}
\usepackage{mathrsfs}
\usepackage{graphicx} 
\usepackage{enumitem}
\usepackage{authblk}
\usepackage[ruled,linesnumbered]{algorithm2e}  
\usepackage[top=2cm, bottom=2cm, left=2.5cm, right=2.5cm]{geometry} 

\title{Modeling Coupled Epidemic-Information Dynamics via Reaction-Diffusion Processes on Multiplex Networks with Media and Mobility Effects}
\author{Guangyuan Mei, Yao Cai, Su-Su Zhang, Ying Huang, Chuang Liu\thanks{Chuang Liu is with the Research Center for Complexity Sciences, Hangzhou Normal University, Hangzhou 311121, China. \texttt{liuchuang@hznu.edu.cn}}, Xiu-Xiu Zhan\thanks{Xiu-Xiu Zhan is with the Research Center for Complexity Sciences, Hangzhou Normal University, Hangzhou 311121, China, and the College of Media and International Culture, Zhejiang University, Hangzhou 310058, China. \texttt{zhanxiuxiu@hznu.edu.cn }}}

\begin{document}
\maketitle
\begin{abstract}
    While most existing epidemic models focus on the influence of isolated factors, infectious disease transmission is inherently shaped by the complex interplay of multiple interacting elements. To better capture real-world dynamics, it is essential to develop epidemic models that incorporate diverse, realistic factors. In this study, we propose a coupled disease–information spreading model on multiplex networks that simultaneously accounts for three critical dimensions: media influence, higher-order interactions, and population mobility. This integrated framework enables a systematic analysis of synergistic spreading mechanisms under practical constraints and facilitates the exploration of effective epidemic containment strategies. We employ a microscopic Markov chain approach (MMCA) to derive the coupled dynamical equations and identify epidemic thresholds, which are then validated through extensive Monte Carlo (MC) simulations. Our results show that both mass media dissemination and higher-order network structures contribute to suppressing disease transmission by enhancing public awareness. However, the containment effect of higher-order interactions weakens as the order of simplices increases. We also explore the influence of subpopulation characteristics, revealing that increasing inter-subpopulation connectivity in a connected metapopulation network leads to lower disease prevalence. Furthermore, guiding individuals to migrate toward less accessible or more isolated subpopulations is shown to effectively mitigate epidemic spread. These findings offer valuable insights for designing targeted and adaptive intervention strategies in complex epidemic settings.
\end{abstract}
\section{Introduction}
The outbreak of infectious diseases has long posed significant threats to public health, social stability, and economic development. As such, accurately modeling epidemic transmission dynamics and devising practical prevention and containment strategies have become central challenges in public health research and management \cite{keeling2005networks,lu2021stability,wang2017unification,li2021dynamics,basnarkov2021seair,boccaletti2020modeling,morris2021optimal}. Global experiences in epidemic control have shown that the development of vaccines, the dissemination of disease-related information, and the spatial mobility of individuals all play crucial roles in shaping the trajectory of infectious disease spread. Prior to the availability of effective vaccines, enhancing public awareness through information diffusion and limiting the spatial spread of infections remain among the most effective measures for mitigating outbreaks.

During an epidemic, related information often spreads rapidly through online social networks \cite{zhan2018coupling,anwar2020role,zhang2023study,wu2024influence}. In this context, the propagation of the disease accelerates the diffusion of information, while information dissemination simultaneously influences individual behaviors-forming a feedback loop that further affects the epidemic’s dynamics \cite{kabir2019analysis,xu2015epidemic,funk2009spread}. These interactions constitute a coupled nonlinear system. In recent years, researchers have developed epidemic–information co-spreading models using multiplex networks \cite{han2024impact,huang2021modeling,kabir2019effect}, in which nodes across layers represent the same individuals in different functional dimensions. For instance, the Unaware–Aware–Unaware–Susceptible–Infected–Susceptible (UAU–SIS) model \cite{granell2013dynamical,hong2022co} uses a UAU process to simulate information spread on the virtual layer and an SIS process to model disease transmission on the physical contact layer, with both types of spreading typically occurring via pairwise (dyadic) interactions.

However, in real-world social systems, information is not transmitted solely through binary interactions. Group communication mechanisms and media influence also play key roles. For example, on platforms like Twitter, users receive information not only through direct contacts but also via group interactions (e.g., group chats, retweets) and from mass media accounts they follow \cite{xia2024dynamic,pei2015exploring,zhao2014fluxflow,li2025effect}. Such non-dyadic influences can be effectively captured through higher-order structures such as simplicial complexes, which allow for a more realistic representation of group-based information diffusion.

In contrast, the spatial spread of infectious diseases is often driven by the mobility of individuals between geographic regions via transportation systems. Individual movement thus constitutes another key factor in epidemic dynamics \cite{soriano2018spreading,keeling2010individual,feng2020epidemic}. To model these dynamics, reaction–diffusion–based metapopulation models have been widely adopted \cite{gao2022epidemic,shao2022epidemic,soriano2018spreading}. These models divide the population into subpopulations (e.g., communities, cities, countries), with connections between them representing transportation links. Disease spreads locally within subpopulations through interpersonal contact and propagates across regions through individual migration. By adjusting the structure of subpopulations and inter-regional mobility, such models can evaluate the impact of mobility and containment measures (e.g., lockdowns) on the course of an epidemic.

Despite considerable progress, existing models still face several limitations in capturing the full complexity of real-world epidemic spreading. First, although some studies have examined the influence of higher-order structures on disease or information propagation, the role of their order and prevalence in coupled spreading dynamics remains unclear. Second, while population mobility is known to significantly influence disease transmission, its interaction with simultaneous information diffusion is not fully understood. Third, most existing epidemic–information co-spreading models oversimplify or neglect important real-world features, such as media influence, mobility patterns, and group-level interactions, limiting their ability to reflect actual transmission processes \cite{kar2019stability,gao2022epidemic,zhu2023epidemic,granell2013dynamical,ruan2012epidemic,benson2016higher,nie2023voluntary}.

To overcome these limitations, we propose a novel coupled disease–information dynamic model that integrates media influence, higher-order interactions, and population mobility within a multiplex network framework. In our model, the virtual information layer incorporates both higher-order simplicial structures and media effects, while the physical contact layer is modeled as a metapopulation network governed by a reaction–diffusion process. This design enables a more faithful representation of real-world epidemic transmission, facilitating the discovery of emergent dynamics, the identification of key influencing factors, and the development of more effective intervention strategies.

The main contributions of this work are summarized as follows:
\begin{itemize}
\item We propose a coupled spreading model that integrates metapopulation mobility and mass media influence. The model is constructed on a two-layer network: a physical contact layer where disease spreads, and a virtual information layer where awareness propagates. A designated media node is introduced in the information layer to capture real-world media effects.

\item We analyze how higher-order group interactions and media broadcasting influence information diffusion and indirectly affect epidemic outcomes. Our results show that both mechanisms enhance public awareness, which in turn helps mitigate disease transmission. This suggests that strengthening media campaigns and leveraging group-based communication can be effective for epidemic control.

\item By integrating metapopulation dynamics, we reveal nuanced impacts of mobility on epidemic transmission. Increasing connectivity within a BA-structured connected metapopulation network, i.e., by adding edges, can reduce overall disease prevalence. Moreover, moderate levels of migration toward sparsely connected regions significantly suppress epidemic spread. These findings offer fresh perspectives on mobility-informed strategies for epidemic control.

\end{itemize}
The remainder of this paper is organized as follows. Section 2 introduces the proposed model in detail. Section 3 presents the coupled MMCA equations and analyzes the epidemic threshold. Section 4 reports the simulation results and investigates the influence of key factors. Section 5 concludes with a summary of findings and discussions on practical implications for epidemic containment.

\section{Model}
To explore the underlying mechanisms of infectious disease transmission under various real-world factors, we develop a coupled disease-information propagation model based on multiplex networks, which includes metapopulation mobility and the impact of mass media. The model features a two-layer network and a media node responsible for disseminating information. Table \ref{table:1} provides a summary of the mathematical notations used in this study. We will now present a detailed introduction to our proposed model and its propagation dynamics.

\begin{table}
\caption{Symbol explanations} 
\label{table:1}
\centering
\begin{tabular}{>{\centering\arraybackslash}m{1.5cm}|>{\raggedright\arraybackslash}m{13cm}}
\toprule 
Paramters   & Descriptions
\\
\midrule 
\rowcolor[HTML]{EFEFEF} 
\textbf{$N$}    & Number of nodes. 
\\ 
\textbf{$M$}    & Number of subpopulations. 
\\ 
\rowcolor[HTML]{EFEFEF} 
\textbf{$\beta _{d}$}& Probability of an unaware node being informed in a simplex of order $d$.\\
\textbf{$\gamma $}    & Probability that aware nodes forget information and become unaware.
\\ 
\rowcolor[HTML]{EFEFEF} 
\textbf{$\lambda$}    & Probability that a susceptible node becomes infected through a single contact with an infected node.
\\
\textbf{$\alpha $}    & Attenuation parameters for the probability of AS (aware and susceptible) nodes being infected through a single contact with an infected node.
\\
\rowcolor[HTML]{EFEFEF} 
\textbf{$\mu$}    & Recovery rate of an infected node.
\\ 
\textbf{$\omega$}    & Probability of an unaware node being informed through the media and becoming aware.\\
\rowcolor[HTML]{EFEFEF} 
\textbf{$\sigma $}    & Enhanced parameter for the probability of UI (unaware and infected) nodes being informed through the media and becoming aware.
\\
\textbf{$g^{U}$}    & The migration probability of unaware nodes.
\\ 
\rowcolor[HTML]{EFEFEF} 
\textbf{$g^{A}$}    & The migration probability of aware nodes.
\\
\textbf{$\Delta $}    & Relation parameter between $g^{U}$ and $g^{A}$, $g^{A}=\Delta g^{U}$.
\\ 
\rowcolor[HTML]{EFEFEF} 
\textbf{$R_{lk}^{U} (t)$}    &  Probability that an unaware node migrates from subpopulation $l$ to subpopulation $k$ at time step $t$.
\\
\textbf{$R_{lk}^{A} (t) $}    & Probability that an aware node migrates from subpopulation $l$ to subpopulation $k$ at time step $t$.
\\ 
\rowcolor[HTML]{EFEFEF} 
\textbf{$b_{lk}$}    &  Elements of the adjacency matrix $B$ representing connections in the metapopulation network.
\\
\textbf{$P_{i}^{X}(t)$}    & Probability that node $i$ is in state $X$ at time step $t$.
\\
\rowcolor[HTML]{EFEFEF} 
\textbf{$C_{i}(t)$}    & Probability that an unaware node $i$ becomes aware at time step $t$ through the non-media information diffusion.
\\ 
\textbf{$\mathscr{S}_{i}$}& The set of all non-zero order simplex including node $i$.
\\
\rowcolor[HTML]{EFEFEF} 
\textbf{$m_{i}^{S}(t)$}    & Probability of an unaware susceptible node $i$ becoming aware at step $t$.
\\ 
\textbf{$m_{i}^{I}(t)$}    & Probability of an unaware infected node $i$ becoming aware at step $t$.
\\
\rowcolor[HTML]{EFEFEF} 
\textbf{$E_{k\to l} (t)$}    & Expected number of nodes migrating from subpopulation $k$ to subpopulation $l$ at time step $t$.
\\
\textbf{$q_{l}^{U}(t)$}    & Probability that an unaware susceptible node in subpopulation $l$ becomes infected after the individual movement stage at time step $t$.\\ 
\rowcolor[HTML]{EFEFEF} 
\textbf{$q_{l}^{A}(t)$}    & Probability of an aware susceptible node in subpopulation $l$ becoming infected after the individual movement stage at time step $t$.\\
\textbf{$\rho ^{X}$}    & Proportion of nodes in state $X$ at steady state.
\\ 
\rowcolor[HTML]{EFEFEF} 
\textbf{$\rho _{k}^{I}(t)$}    & Proportion of infected nodes in subpopulation $k$ at time step $t$.
\\
\textbf{$Z_{k}$}    & Number of nodes in subpopulation $k$.
\\ 
\rowcolor[HTML]{EFEFEF} 
\textbf{$D_{k}$}    & Degree of subpopulation $k$.
\\
\textbf{$Q_{l}^{U}(t)$}    & Probability that an unaware susceptible node in subpopulation $l$ becomes infected after the individual return stage at time step $t$.\\ 
\rowcolor[HTML]{EFEFEF}
\textbf{$Q_{l}^{A}(t)$}    & Probability that an aware susceptible node in subpopulation $l$ becomes infected after the individual return stage at time step $t$.\\ \bottomrule 
\end{tabular}
\end{table}
\subsection{Underlying Multiplex Network}
\begin{figure}
    \centering
    \includegraphics[width=0.5\linewidth]{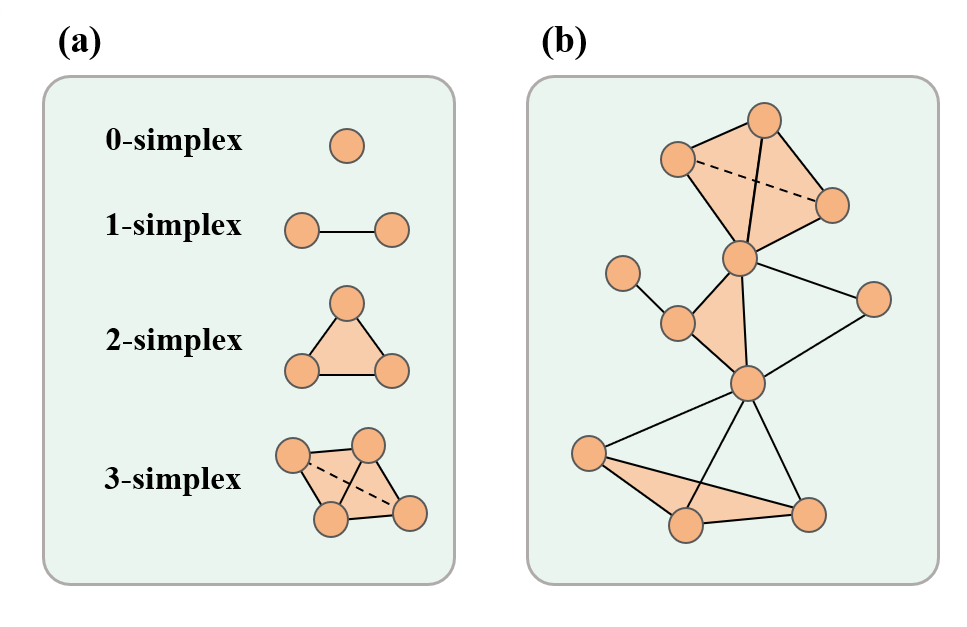}
    \caption{Schematic representation of simplicial complex structures. (a) Illustration of simplices of orders 0-3. (b) A toy example of a simplicial complex network.}
    \label{fig:simplex&net}
\end{figure}

In real-world scenarios, both information diffusion in the virtual layer and disease transmission in the physical layer are shaped not only by pairwise interactions between individuals but also by higher-order group interactions~\cite{ma2024impact,wan2022multilayer,battiston2021physics}. To model these processes in both layers, we assume that the two layers share the same nodes but exhibit different network structures, with the total number of nodes denoted as $N$. To accurately capture these complex, multi-scale interactions within populations, we model the underlying network for information diffusion using a simplicial complex network~\cite{iacopini2019simplicial}. Unlike traditional pairwise networks, simplicial complexes inherently account for group interactions by incorporating higher-dimensional structures called simplices.   A simplicial complex consists of simplices of various orders, where a $d$-dimensional simplex ($d$-simplex) includes $d+1$ nodes, with all pairwise edges connecting them.
As shown in Figure \ref{fig:simplex&net}(a), a
0-simplex represents a single node, a 1-simplex corresponds to an edge between two nodes, a 2-simplex forms a triangular face linking three nodes, and a 3-simplex extends into a tetrahedral structure with four nodes interconnected by six edges. Importantly, each 
$d$-simplex is composed of multiple lower-order simplices. For example, a 2-simplex contains three 1-simplices and three 0-simplices, while a 3-simplex consists of four 2-simplices, six 1-simplices, and four 0-simplices. Figure \ref{fig:simplex&net}(b) illustrates an example of a small-scale simplicial complex network, highlighting the integration of higher-order interactions within a networked system.

The physical layer is modeled using a metapopulation framework to capture node mobility. Specifically, $N$ nodes are randomly distributed across $M$ subpopulations, and edges between these subpopulations form a transportation network that facilitates individual mobility. As a result, in the proposed model, infectious diseases can spread not only within individual subpopulations but also between subpopulations through the movement of infected nodes.

We show an example of the two layer network in Figure \ref{fig:model&spread}(a), i.e., an upper virtual information layer, represented by simplices, to simulate information diffusion within the population, and a lower physical contact layer, structured as a metapopulation, to model disease transmission. Furthermore, to capture the influence of external factors such as government or media, we introduce a media node that connects to all nodes in the virtual information layer.

\subsection{Modeling the Coupling Dynamics between Disease and Information}

The process of information diffusion follows the (UAU) model, where each node exists in one of two possible states: Aware (A) or Unaware (U). An unaware node can transition to the aware state by acquiring information about the disease through interactions with its neighbors, within groups, or via mass media exposure. To quantify this process, we define $\beta_d$ as the probability that an unaware node becomes aware through a $d$-order simplex interaction ($d=1, 2$ or $3$), capturing direct pairwise interactions, small group influences, and higher-order collective effects, respectively. Additionally, $\omega$ denotes the probability that an unaware node transitions to the aware state due to mass media influence. The evolution of information diffusion in this model is governed by the following state transition dynamics:

\begin{equation}
\left\{
\begin{aligned}
U + A &\overset{\beta_{d}}{\rightarrow} A + A, \\
U &\overset{\omega}{\rightarrow} A,\\
A &\overset{\gamma}{\rightarrow} U.
\end{aligned}
\right.
\label{formula:1}
\end{equation}

In the physical contact layer, the Susceptible-Infected-Susceptible (SIS) model is used to describe the spread of infectious diseases within a population. Each node exists in one of two states: Susceptible (S) or Infected (I). A susceptible node, upon interacting with an infected neighbor, has a probability $\lambda$ of becoming infected. Infected nodes, in turn, recover and revert to the susceptible state at a rate $\mu$. The propagation mechanism of the SIS model is as follows:
\begin{equation}
\left\{
\begin{aligned}
S + I &\overset{\lambda }{\rightarrow} I + I, \\
I &\overset{\mu }{\rightarrow} S.
\end{aligned}
\right.
\label{formula:2}
\end{equation}
\begin{figure}
    \centering
    \includegraphics[width=1.0\linewidth]{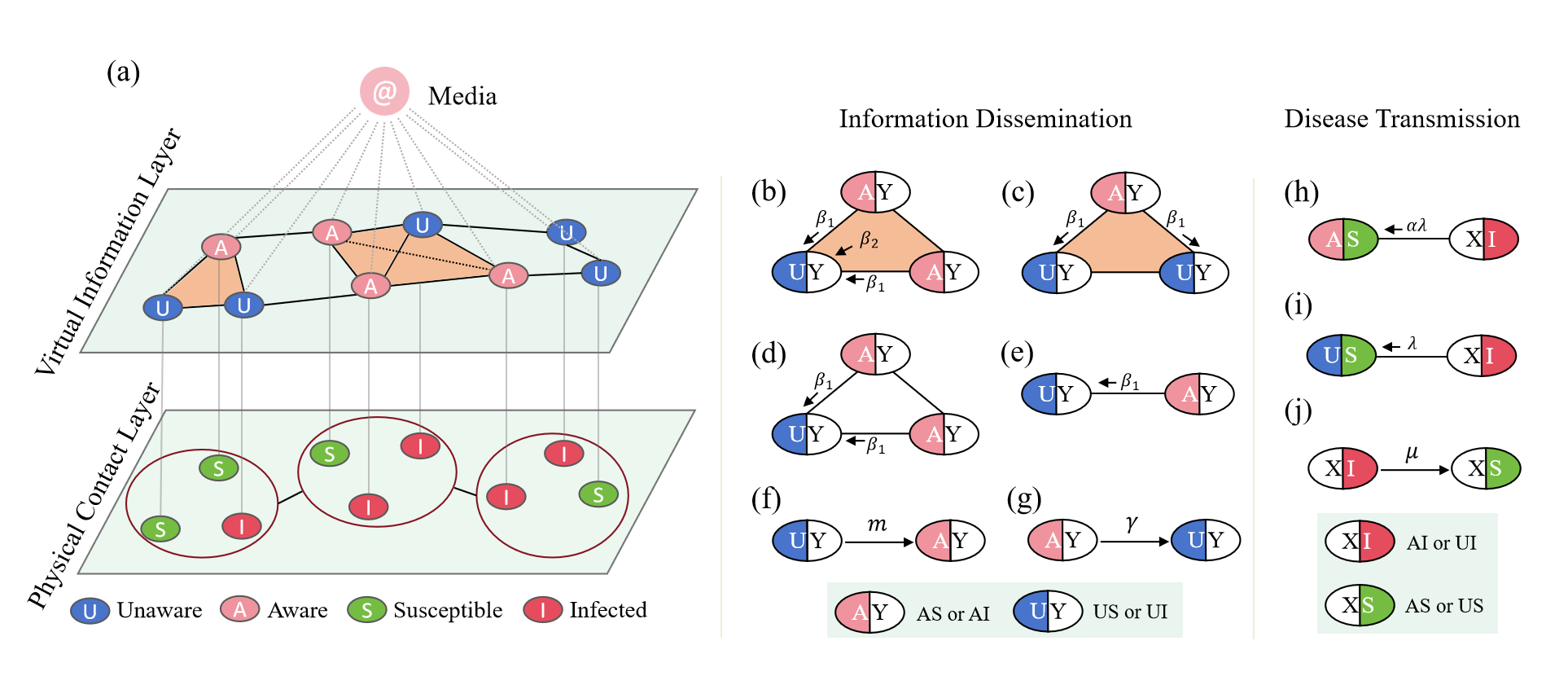}
    \caption{Schematic representation of the proposed model. (a) The underlying network structure of the spreading model. (b-g) The information diffusion process: AY (UY) represents an aware (unaware) state, where the infection status of the node is unspecified (denoted by Y). (b-c) Information transmission through 2-simplices. (d-e) Information transmission through 1-simplex. (f) Transition of an unaware (U) node to an aware (A) state. (g) Transition of an aware (A) node back to an unaware (U) state due to information forgetting. (h-j) The infectious disease spreading process: XI (XS) indicates that a node is in an infected (susceptible) state, where the information status of the node is unspecified (denoted by X).  (h–i) The transmission mechanism of the infectious disease. (j) Recovery of an infected (I) node to the susceptible (S) state.}
    \label{fig:model&spread}
\end{figure}

By integrating the two spreading processes, nodes in the coupled dynamics can be categorized into four distinct states: unaware-susceptible (US) nodes, who have neither information about the disease nor an active infection; unaware-infected (UI) nodes, who are unaware of the disease but are currently infected; aware-susceptible (AS) nodes, who have knowledge of the disease but remain uninfected; and aware-infected (AI) nodes, who are both informed and infected. 
These four states, along with individual mobility patterns, divide the spreading process into four stages at each time step: the information diffusion process, individual movement, disease transmission, and individual return stages. In this model, nodes that receive information during the information dissemination stage instantly become aware of the disease. As a result, these nodes exhibit a lower likelihood of movement in the movement stage and are at a reduced risk of infection during the infectious disease transmission stage. We will now explore the specifics of each of these four coupled transmission stages in detail. In the following, we show the details of each stage:

\begin{itemize}
\item \textbf{Information diffusion process.}
Each unaware node (U) becomes aware with probability $\omega$ through mass media. As shown in Figure \ref{fig:model&spread}(b), if a U-state node is part of a $d$-simplex and $d$ nodes within this simplex are in the A-state, the joint effect of these $d$ nodes causes the U node to become aware with probability $\beta_{d}$. Additionally, aware nodes (A) may forget the disease-related information and revert to the unaware state with probability $\gamma$, as shown in Figure \ref{fig:model&spread}(g).

\item \textbf{Individual movement.}
This stage describes the movement of nodes between subpopulations. A U-state node moves to neighboring subpopulations with probability $g^{U}$. A-state nodes, due to their protective awareness, exhibit reduced mobility and move with a lower probability, $g^{A}=\Delta g^{U}$, where $0<\Delta<1$. After determining whether to move, nodes select a target subpopulation. The probability of a U-state node from subpopulation $l$ moving to subpopulation $k$ is given by: 

\begin{equation} R_{lk}^{U}=\frac{b_{lk} }{\sum_{m=1}^{M}b_{lm}}, \label{formula:3} 
\end{equation}

where $M$ is the number of subpopulations, and $b_{lk}$ and $b_{lm}$ are elements of the adjacency matrix $B$ of the metapopulation network. Due to their protective awareness, A-state nodes preferentially move to subpopulations with a lower proportion of infected nodes. Therefore, the probability of an A-state node from subpopulation $l$ moving to subpopulation $k$ is: 

\begin{equation} R_{lk}^{A}(t)=\frac{b_{lk}\mathcal{S}_{k}(t)}{\sum{m=1}^{M}b_{lm}\mathcal{S}_{m}(t)},
\label{formula:4}
\end{equation} 

where $\mathcal{S}_{k}(t)$ represents the number of S nodes in subpopulation $k$ at time step $t$. At the end of the movement stage, some nodes remain in their initial subpopulations, while others migrate to neighboring subpopulations.

\item \textbf{Disease spreading process.}
This stage describes the transmission of the infectious disease within the subpopulation. As shown in Figure \ref{fig:model&spread}(i), infected nodes (I) spread the disease to all US-state nodes in their subpopulation with probability $\lambda$. Due to protective measures, the probability of AS-state nodes becoming infected is reduced to $\alpha \lambda$, where $0<\alpha<1$ (Figure \ref{fig:model&spread}(h)). I-state nodes recover to the S state with probability $\mu$, as depicted in Figure \ref{fig:model&spread}(j). Unlike traditional coupled information-disease models, our approach includes nodes in the UI state, i.e., infected nodes may obtain disease-related information and then transition to the aware-infected state with probability $\sigma \omega$, where $\sigma > 1$.

\item \textbf{Individual return stage.}
This stage involves the return of nodes that have migrated back to their original subpopulations, which are the subpopulations they were part of before the coupled spreading process started. After the mobility phase, some nodes move away from their initial subpopulations, and during the return stage, these nodes rejoin their original subpopulations.
\end{itemize}

\section{Theoretical Analysis Based on Microscopic Markov Chain Approach}
MMCA models the state evolution of network nodes as a Markov process, examining network dynamics through local topological structures and state transition probabilities. It assumes that a node's state at time step $t$ is determined by its own state and the states of its neighboring nodes at the previous time step $t-1$ \cite{zhan2018coupling,soriano2018spreading,gao2022epidemic}. In this section, we analyze the dynamical equations of the proposed model and derive the epidemic transmission threshold by MMCA.


\subsection{Microscopic Markov Chain Equations}
Based on the model described in the previous section, we derive the state transition probability for each node. Let $P_{i}^{X} (t)$ represent the probability that node $i$ is in state $X$ at time $t$, where $X \in \{U, A, S, I, US, UI, AS, AI\}$. The following equation then holds:
\begin{equation}
\left\{\begin{aligned}
P_{i}^{U} (t)&=P_{i}^{US} (t)+P_{i}^{UI} (t),
 \\P_{i}^{A} (t)&=P_{i}^{AS} (t)+P_{i}^{AI} (t),
 \\P_{i}^{S} (t)&=P_{i}^{US} (t)+P_{i}^{AS} (t),
 \\P_{i}^{I} (t)&=P_{i}^{UI} (t)+P_{i}^{AI} (t).
\end{aligned}\right.
\label{formula:5}
\end{equation}
A U-state node can become informed either through mass media or by interacting with its neighbors. Let $C_{i} (t)$ denote the probability that a U-state node $i$ transitions to the A state at time step $t$ due to interactions with neighboring nodes. Thus, we have:
\begin{equation}
\left\{
\begin{aligned}
C_{i} (t) &= 1-\left \{ \prod_{\mathscr{S} _{g}^{k}\in \mathscr{S} _{i}} \left [ 1-\beta _{k} \prod_{j\ne i,\, j\in \mathscr{S} _{g}^{k}} P_{j}^{A}(t)  \right ]  \right \} , \\
\mathscr{S}_{i} &= \left ( \mathscr{S} _{1}^{k_{p} },\mathscr{S} _{2}^{k_{l} },\cdots ,\mathscr{S} _{x_{i} }^{k_{e} } \right ),
\end{aligned}
\right.
\label{formula:6}
\end{equation}
where $\mathscr{S}_{i}$ denotes the set of all nonzero-order simplices containing node $i$, with each element $\mathscr{S}_{g}^{k}$ representing a simplex of order $k$. 
The cardinality of $\mathscr{S}_{i}$ is given by $x_i$. Let $m_{i}^{S} (t)$ and $m_{j}^{A} (t)$  denote the probabilities that a US-state node $i$ and a UI-state node $j$ , respectively, transition to the aware state at time $t$. These probabilities are determined by the following equations:

\begin{equation}
\left\{
\begin{aligned}
m_{i}^{S} (t)&=1-\left \{ \left ( 1-\omega  \right ) \left [ 1-C_{i}(t)  \right ]  \right \},\\
m_{i}^{I} (t)&= 1-\left \{ \left ( 1-\sigma\omega  \right ) \left [ 1-C_{i}(t)  \right ]  \right \}.
\end{aligned}
\right.
\label{formula:7}
\end{equation}

To quantify individual movement as described in the previous section, we denote $E_{k \rightarrow l}(t)$ as the expected number of nodes migrating from subpopulation $k$ to subpopulation $l$ at time $t$. The calculation formula is as follows:
\begin{equation}
\begin{aligned}
 E_{k \rightarrow l}(t) &= \delta_{kl} \left\{  \sum_{j=1}^{Z_{k} } P_j^A(t)(1-\gamma)(1-g^A) 
     + \sum_{j=1}^{Z_{k} } P_j^U(t)\left [ P_j^S(t)m_{j}^{S}(t)+P_j^I(t)m_{j}^{I}(t) \right ] (1-g^A)\right. \\
& \hphantom{\sigma_{kl}}\left.  + \sum_{j=1}^{Z_{k} } P_j^A(t)\gamma(1-g^U)
    + \sum_{j=1}^{Z_{k} } P_j^U(t) \left[ 1-P_j^S(t)m_{j}^{S}(t)-P_j^I(t)m_{j}^{I}(t)\right](1-g^U) \right\} \\
&\hphantom{\sigma_{kl}} + \sum_{j=1}^{Z_{k} } P_j^A(t)(1-\gamma)g^A R_{kl}^A 
 + \sum_{j=1}^{Z_{k} } P_j^U(t)\left[ 1-P_j^S(t)m_{j}^{S}(t)-P_j^I(t)m_{j}^{I}(t)\right]g^U R_{kl}^U\\
&\hphantom{\sigma_{kl}} + \sum_{j=1}^{Z_{k} } P_j^A(t)\gamma g^U R_{kl}^U+ \sum_{j=1}^{Z_{k} } P_j^U(t)\left [ P_j^S(t)m_{j}^{S}(t)+P_j^I(t)m_{j}^{I}(t) \right ]g^A R_{kl}^A ,
\end{aligned}
\label{formula:8}
\end{equation}
where $\delta_{kl}$ is the Kronecker delta function, defined as:
\begin{equation}
\delta_{kl} =
\begin{cases}
1 & if \quad  k = l, \\
0 & otherwise.
\end{cases}
\label{formula:9}
\end{equation}

\begin{figure}
    \centering
    \includegraphics[width=0.7\linewidth]{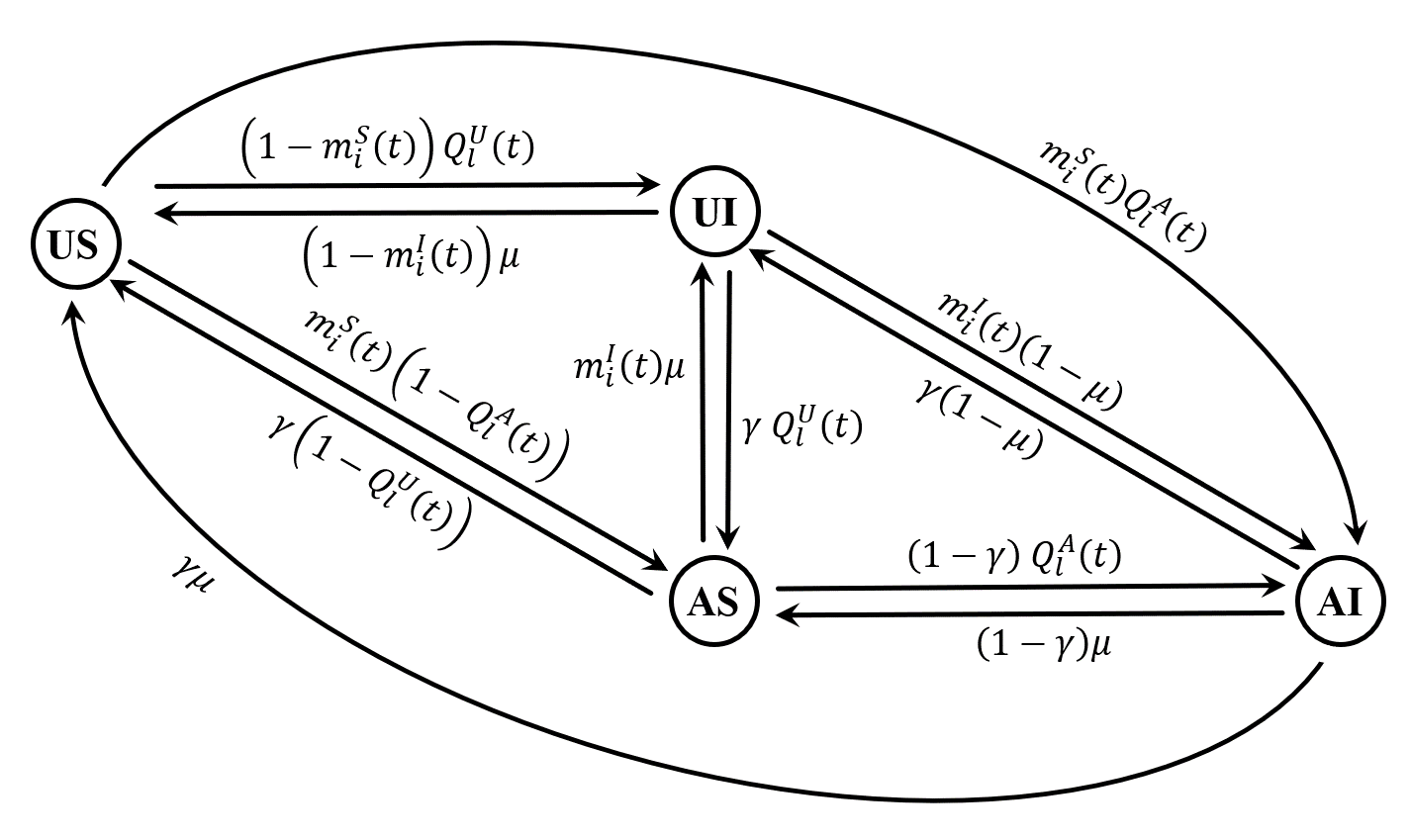}
    \caption{State transition diagram of node $i$ in subpopulation $l$. The symbols along the arrows indicate the transition probabilities between different states.
}
    \label{fig:state}
\end{figure}

Let $q_{l}^{U} (t)$ and $q_{l}^{A} (t)$ denote the probabilities that US-state nodes and AS-state nodes in subpopulation $l$ become infected after the movement phase at time step $t$, respectively. These probabilities are computed as follows:
\begin{equation}
\left\{
\begin{aligned}
q_{l}^{U} (t)&=1-\prod_{k}^{} \left [ 1-  \lambda \rho _{k}^{I}(t)\right ] ^{E_{k \rightarrow l}(t)} ,\\
q_{l}^{A} (t)&=1-\prod_{k}^{} \left [ 1-\alpha \lambda \rho _{k}^{I}(t) \right ] ^{E_{k \rightarrow l}(t)},
\end{aligned}
\right.
\label{formula:10}
\end{equation}
where $\rho _{k}^{I}(t)=\left [ \sum_{i\in k }^{} P_{i}^{I} \left ( t \right )  \right ] /Z_{k}$ represents the proportion of I-state nodes in subpopulation $k$ at time $t$, with $Z_{k}$ denoting the total number of nodes in subpopulation $k$. Let $Q_{l}^{U}\left (t \right )$ and $Q_{l}^{A}\left (t \right )$ represent the probabilities that US-state nodes and AS-state nodes in subpopulation $l$ become infected after the individual return stage at time $t$, respectively. It is worth noting that both $q_{l}^{X}(t)$ and $Q_{l}^{X}(t)$ describe the probability of XS-state nodes (where X = A \ or\ U) in subpopulation $l$ being infected. However, they correspond to different stages of the transmission process within time $t$. The probability $Q_{l}^{X}(t)$ is given by:
\begin{equation}
\left\{
\begin{aligned}
Q_{l}^{U} (t)&=g^{A}\sum_{k=1,k\ne l}^{M}R_{lk}^{A}(t)q_{k}^{A}(t)+\left ( 1-g^{A} \right ) q_{l}^{A}(t),\\
Q_{l}^{A} (t)&=g^{U}\sum_{k=1,k\ne l}^{M}R_{lk}^{U}(t)q_{k}^{U}(t)+\left ( 1-g^{U} \right ) q_{l}^{U}(t).
\end{aligned}
\right.
\label{formula:11}
\end{equation}
The expression $Q_{l}^{X}(t)$, where $X=\{U, A\}$, consists of two distinct terms. The first term quantifies the probability that S-state nodes originating from subpopulation $l$ migrate to other subpopulations and acquire infection during their movement. The second term captures the probability that 
S-state nodes remain within subpopulation $l $ without migrating and become infected locally.
By integrating Eq. (\ref{formula:7}) and Eq. (\ref{formula:11}), we can derive the transition probabilities governing the state changes of nodes, as depicted in Figure \ref{fig:state}.

Based on the transition probabilities between different states of nodes in subpopulations, we can determine the likelihood that a node is in a specific state at the next time step. This can be expressed as follows:
\begin{align}
\left\{
\begin{array}{l}
P_i^{US}(t+1) = \left(1 - m_i^S(t)\right) \left(1 - Q_l^U(t)\right) P_i^{US}(t) + \gamma \left(1 - Q_l^U(t)\right) P_i^{AS}(t) \\
\quad \quad \quad \quad \quad \quad+ (1 - m_i^I(t)) \mu P_i^{UI}(t) + \gamma \mu P_i^{AI}(t), \\
P_i^{UI}(t+1) = \left(1 - m_i^I(t)\right) (1 - \mu) P_i^{UI}(t) + \gamma (1 - \mu) P_i^{AI}(t) \\
\quad \quad \quad \quad \quad \quad + \left(1 - m_i^S(t)\right) Q_l^U(t) P_i^{US}(t) + \gamma Q_l^U(t) P_i^{AS}(t), \\
P_i^{AS}(t+1) = (1 - \gamma) \left( (1 - Q_l^A(t)\right) P_i^{AS}(t) + m_i^S(t)\left(1 - Q_l^A(t)\right) P_i^{US}(t) \\
\quad \quad \quad \quad \quad \quad+ (1 - \gamma) \mu P_i^{AI}(t) + m_i^I(t) \mu P_i^{UI}(t), \\
P_i^{AI}(t+1) = (1 - \gamma)(1 - \mu) P_i^{AI}(t) + m_i^I(t)(1 - \mu) P_i^{UI}(t) \\
\quad \quad \quad \quad \quad \quad + (1 - \gamma) Q_l^A(t) P_i^{AS}(t) + m_i^S(t) Q_l^A(t) P_i^{US}(t).
\end{array}
\right.\label{formula:12}
\end{align}

\subsection{Mathematical Derivation of the Epidemic Transmission Threshold}

The transmission threshold $\lambda_{c}$ plays a critical role in determining whether an infectious disease is capable of spreading through a population. If the transmission probability $\lambda$ exceeds the threshold $\lambda_{c}$, the disease can propagate with the population; otherwise, it will not. As time progresses and $t\to \infty$, the system stabilizes, and the proportion of individuals in each state reaches equilibrium, meaning that the state of each individual no longer changes, i.e., $P_{i}^{X} (t+1)=P_{i}^{X} (t)$. Consequently, Eq. (\ref{formula:12}) can be written as:

\begin{align}
\left\{
\begin{array}{l}
P_i^{US} = \left(1 - m_i^S\right) \left(1 - Q_l^U\right) P_i^{US} + \gamma \left(1 - Q_l^U\right) P_i^{AS}  + (1 - m_i^I) \mu P_i^{UI} + \gamma \mu P_i^{AI},\\ 
P_i^{UI} = \left(1 - m_i^I\right) (1 - \mu) P_i^{UI} + \gamma (1 - \mu) P_i^{AI} + \left(1 - m_i^S\right) Q_l^U P_i^{US} + \gamma Q_l^U P_i^{AS},\\ 
P_i^{AS} = (1 - \gamma) \left( 1 - Q_l^A\right) P_i^{AS} + m_i^S\left(1 - Q_l^A\right) P_i^{US} + (1 - \gamma) \mu P_i^{AI} + m_i^I \mu P_i^{UI}, \\ 
P_i^{AI} = (1 - \gamma)(1 - \mu) P_i^{AI} + m_i^I(1 - \mu) P_i^{UI} + (1 - \gamma) Q_l^A P_i^{AS} + m_i^S Q_l^A P_i^{US}.
\end{array}
\right. \label{formula:13}
\end{align}

When $\lambda \approx \lambda _{c}$, the epidemic dynamics within the metapopulation network undergo a critical transition from a disease-free state to an endemic equilibrium. At this threshold, the proportion of infected nodes in the system becomes negligibly small, i.e., $P_{i}^{I}=\varepsilon _{i}^{l} \ll 1$. Consequently, we have $P_{i}^{US}\approx P_{i}^{U},P_{i}^{AS}\approx P_{i}^{A}$, and $P_{i}^{U}=1-P_{i}^{A}$, as derived from Eq. (\ref{formula:5}). Furthermore, by neglecting higher-order terms in Eq. (\ref{formula:10}), we derive the following approximation:
\begin{equation}
\left\{
\begin{aligned}
q_{l}^{U} &\approx \sum_{k=1}^{M}\lambda\varepsilon _{i}^{k}E_{k \rightarrow l},\\
q_{l}^{A} &\approx \sum_{k=1}^{M}\alpha \lambda\varepsilon _{i}^{k}E_{k \rightarrow l}.
\end{aligned}
\right.
\label{formula:14}
\end{equation}

To derive the spreading threshold, we assume that each subpopulation contains an equal number of nodes, i.e., $Z_{k}=N/M$. As $\lambda \rightarrow \lambda _{c}$, the probabilities of U-state nodes in subpopulation $l$ and A-state nodes migrating to a neighboring subpopulation $k$ become approximately equal, i.e., $R_{lk}^{U}\approx R_{lk}^{A}$. By combining Eq. (\ref{formula:11}) with Eq. (\ref{formula:13}), we obtain:
\begin{equation}
Q_{l}^{A}\approx \alpha \Delta Q_{l}^{U} + (1-\Delta)\alpha q_{l}^{U}.\label{formula:15} 
\end{equation}

According to Eq. (\ref{formula:13}), we have:
\begin{equation}
\mu \varepsilon _{i}^{l} = \left [ (1-m_{i}^{S} )P_{i}^{US}+\gamma P_{i}^{AS}  \right ] Q_{l}^{U} +\left [m_{i}^{S}P_{i}^{US}+(1-\gamma )P_{i}^{AS}  \right ]Q_{l}^{A} ,\label{formula:16}
\end{equation}
substituting Eq. (\ref{formula:11}), (\ref{formula:14}) and (\ref{formula:15}) into Eq. (\ref{formula:16}), we derive:
\begin{equation}
\begin{aligned}
\frac{\mu}{\lambda} \varepsilon_{i}^{l}=\sum_{k=1}^{M}\left\{\begin{array}{l}
\left\{\left[1+(\alpha \Delta-1) m_{i}^{s}\right]\left(\mathbf{R_i} -1\right) g^{U}+\left[1+(a-1) m_{i}^{s}\right]\right\} \\
+\left[(\alpha \Delta-1)\left(\mathbf{R_i} -1\right) g^{U}+(a-1)\right]\left(1-\gamma-m_{i}^{s}\right) P_{i}^{A}
\end{array}\right\} E_{k \rightarrow l} \varepsilon_{i}^{l},
\end{aligned}
\label{formula:23}
\end{equation}
where $\mathbf{R_i}$ is an element in  $M\times M$ matrix $\mathbf{R}$, if node $i$ is in subpopulation $l$, then $\mathbf{R_i} = R_{lk}^{U}$. Then, Eq. (\ref{formula:8}) can be expressed as:
\begin{equation}
\begin{aligned}
E_{k \rightarrow l}=\sum_{j\in k}^{}\left\{\begin{array}{c}
{\left[1-g^{U}+(1-\Delta) m_{j}^{S} g^{U}+\left[\left(1-\gamma-m_{j}^{S}\right)(1-\Delta) g^{U}\right] P_{j}^{A}\right] \delta_{k l}} \\
+\left\{\left(\Delta m_{j}^{S}+1-m_{j}^{S}\right)+\left[\left(1-\gamma-m_{j}^{S}\right)(\Delta-1)\right] P_{j}^{A}\right\} g^{U}\mathbf{R_i}  
\end{array}\right\},
\end{aligned}
\label{formula:17}
\end{equation}

Next, by substituting Eq. (\ref{formula:18}) into Eq. (\ref{formula:17}), we obtain:
\begin{equation}
\frac{\mu}{\lambda} \varepsilon_{i}^{l}=\sum_{k=1}^{M} \sum_{j\in k}^{}\underset{H_{ij}}{\underbrace{\left\{\begin{array}{c}
\left\{\begin{array}{c}
\left\{\left[1+(\alpha \Delta-1) m_{i}^{S}\right]\left(\mathbf{R_i}-1\right) g^{U}+\left[1+(a-1) m_{i}^{S}\right]\right\} \\
+\left[(\alpha \Delta-1)\left(\mathbf{R_i}-1\right) g^{U}+(a-1)\right]\left(1-\gamma-m_{i}^{S}\right) P_{i}^{A}
\end{array}\right\}\cdot  \\
\left\{\begin{array}{c}
{\left[1-g^{U}+(1-\Delta) m_{j}^{S} g^{U}+\left(1-\gamma-m_{j}^{S}\right)(1-\Delta) g^{U} P_{j}^{A}\right] \delta_{k l}} \\
+\left\{\left(\Delta m_{j}^{S}+1-m_{j}^{S}\right)+\left[\left(1-\gamma-m_{j}^{S}\right)(\Delta-1)\right] P_{j}^{A}\right\} g^{U} \mathbf{R_i}
\end{array}\right\}
\end{array}\right\}}}  \varepsilon_{i}^{l}.
\label{formula:18}
\end{equation}

Thus, Eq. (\ref{formula:18}) can be formulated as an eigenvalue problem for a feasible solution $\varepsilon_{i}^{l}$. Let $\Lambda _{max}\left \{ H \right \}$ denote the maximum eigenvalue of the matrix $H=\{H_{ij}\}_{N \times N}$. Accordingly, the threshold for the spread of infectious diseases is:
\begin{equation}
\lambda _{c}=\frac{\mu }{\Lambda _{max}\left \{ H \right \}} .
\label{formula:19}
\end{equation}
Numerous studies have explored the influence of various parameters, such as the infectious disease transmission rate, recovery rate, and information forgetting rate, on the transmission threshold \cite{kabir2019analysis,funk2009spread,hong2022co,shao2022epidemic}. In this work, we shift our focus to examining how the introduction of different models or structures, including media, simplices, and the metapopulation framework, impacts the transmission threshold (see Sections \ref{sec:experimental}).
As indicated by Eqs. (\ref{formula:18}) and (\ref{formula:19}), the number of nodes within a subpopulation is a critical factor influencing the coupled spreading process. Thus, we define $\lambda_{t}=\mu/(\frac{N}{M})$ as the scaling factor for the infectious disease transmission rate $\lambda$ in the following experiment, capturing its dependence on subpopulation size.

\section{Experimental results}
\label{sec:experimental}
In this section, we utilize MC simulations to validate the effectiveness of our proposed theoretical framework and examine the impact of media influence, higher-order structures, and population mobility on the coupled information-disease spreading dynamics.
For the network configuration, the virtual information layer is generated using the Barabási-Albert (BA) model~\cite{barabasi1999emergence}. In particular, based on the topological characteristics of simplicial structures, nodes are represented as 0-simplices, edges as 1-simplices, triangular formations as 2-simplices, and tetrahedral formations as 3-simplices, considering simplices up to order 3. Meanwhile, interconnections between subpopulations are also constructed following the BA model, with nodes evenly distributed across subpopulations.


To validate the theoretical MMCA approach, we perform MC simulations. Initially, $1\%$ of the nodes are randomly assigned to the UI state, while the remaining nodes are initialized in the US state. At each time step, the four stages of the coupled spreading process are executed sequentially, with node states updated iteratively until the system stabilizes. The proportion of nodes in state $X$ at steady state, denoted as $\rho ^{X}$, is calculated as:
\begin{equation}
   \rho ^{X}= \frac{1}{N}\sum_{i=1}^{N}P_{i}^{X} , 
\label{formula:20}
\end{equation}
where $X$ represents any of the states $U,S,A$, or $I$.

\begin{figure}
    \centering
    \includegraphics[width=0.6\linewidth]{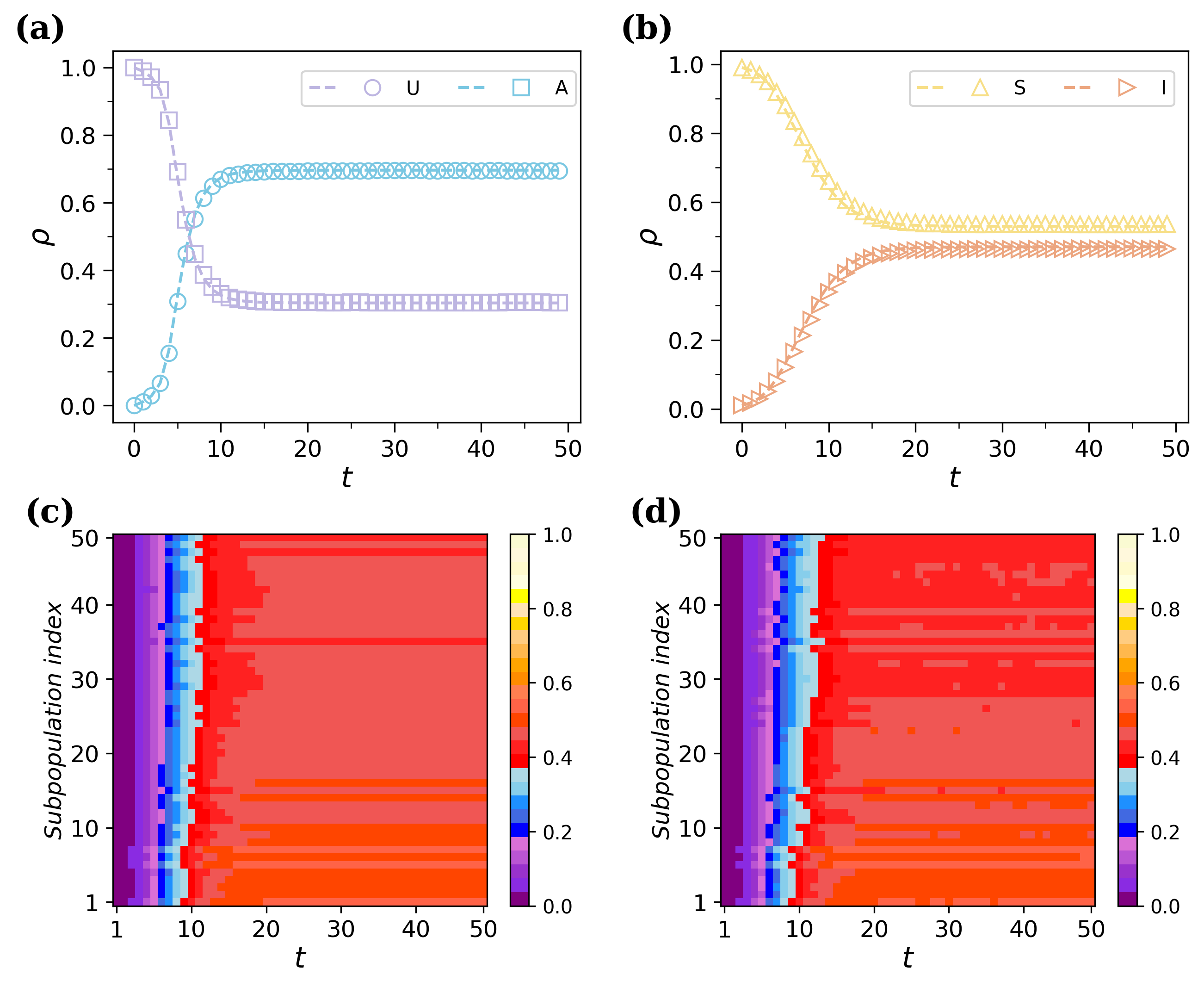}
    \caption{Validation of the MMCA approach through MC simulations. Figures (a) and (b) depict the temporal evolution of the proportion of nodes in different states during the spreading process. The dotted lines represent the theoretical predictions derived from the MMCA approach, while the scatter points correspond to the average results obtained from 200 Monte Carlo simulation runs. 
    Figures (c) and (d) illustrate the variation in the fraction of I-state nodes across subpopulations in the MMCA solutions and MC experiments, respectively. The network consists of $N=5000$ nodes. The virtual information layer is modeled as a BA network with $m = 10$. The metapopulation network in the physical contact layer follows a BA structure with $m=10$, comprising $M=50$ subpopulations, each containing $Z = 100$ nodes. Other parameters are set to $\omega =0.01,\sigma =10,\beta_{1}=0.04, \beta_{2}=0.08, \beta_{3}=0.1, \gamma =0.2,\lambda =0.01,\alpha= 0.5,\mu =0.3,g^{U}=0.6$, and $\Delta =0.5$.}
    \label{fig:nihe}
\end{figure}

We compare the MC simulation results with the MMCA solutions in Figure \ref{fig:nihe}. Figures~\ref{fig:nihe} (a-b) present the temporal evolution of the proportion of nodes in different states ($U, A, S$, and $I$), where dotted lines and scatter points represent MMCA and MC results, respectively. Figures \ref{fig:nihe} (c-d) illustrate the evolution of the I-state proportion using both methods. The strong agreement between MMCA and MC at both the macroscopic (network-wide) and microscopic (subpopulation) levels confirms the validity of the proposed model. Furthermore, Figures \ref{fig:nihe} (a-b) demonstrate that, at the beginning, the prevalence of the disease is low, with only a small fraction of nodes becoming aware through the media, resulting in low proportions of nodes of the A-state and I-state. As interactions continue, the infectious disease spreads, and awareness increases, driving a gradual rise in A-state and I-state proportions. Eventually, due to the compartmental model dynamics, the system stabilizes. Figures \ref{fig:nihe}(c) and (d) show the proportion of I-state nodes in each subpopulation over time. During the transmission process, the proportion of I-state nodes in each subpopulation gradually increases and tends to stabilize.

\begin{figure}
    \centering
    \includegraphics[width=0.7\linewidth]{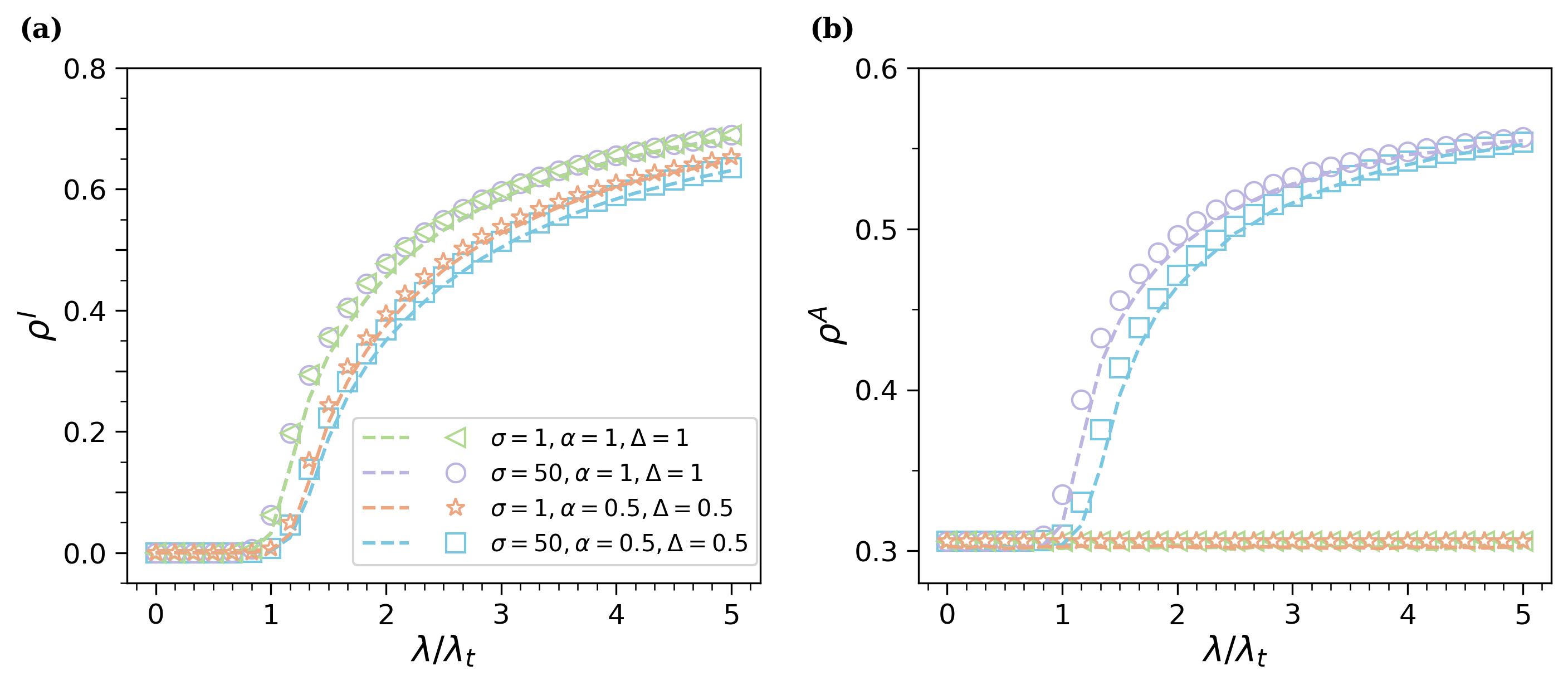}
    \caption{Analysis of the coupling effects between disease transmission and information diffusion. (a) and (b) show the steady-state proportions of I-state nodes ($\rho^{I}$) and A-state nodes ($\rho^{A}$), respectively, as functions of the normalized infection probability $\lambda / \lambda_t$. Different colors represent different combinations of coupling parameters, i.e., $\sigma, \alpha$, and $\Delta$. Dotted lines correspond to theoretical predictions obtained using the MMCA approach, while scatter points represent the average results from 200 MC simulation runs. The virtual layer follows a BA model with $m=10$ and $N=5000$, while the physical layer consists of $M=50$ subpopulations, each containing $Z=100$ nodes, with inter-subpopulation connections also generated using a BA model with $m=10$. Other parameters are set as $\omega =0.01,\beta_{1}=0.04, \beta_{2}=0.08, \beta_{3}=0.1, \gamma =0.2,\lambda =0.01,\mu =0.3$, and $g^{U}=0.6$.}
    \label{fig:couple}
\end{figure}

To explore the interplay between disease transmission and information diffusion, we examine the effects of key coupling parameters-$\sigma$, $\alpha$, and $\Delta$-as shown in Figure~\ref{fig:couple}. Four experimental scenarios are constructed by systematically varying these parameters to simulate different levels of interaction strength between the two spreading processes. The parameter $\sigma$ captures the extent to which disease prevalence promotes information dissemination. A low value ($\sigma = 1$) implies that disease dynamics have no influence on information spread, while a high value ($\sigma = 50$) reflects strong coupling, where increased infection levels significantly drive awareness. Conversely, $\alpha$ and $\Delta$ characterize the reverse effect, i.e., the impact of awareness on disease transmission. When $\alpha = 1$ and $\Delta = 1$, information has no bearing on the epidemic; when $\alpha = 0.5$ and $\Delta = 0.5$, information exerts a strong mitigating effect on disease spread. Figures~\ref{fig:couple}(a) and~\ref{fig:couple}(b) present the steady-state proportions of infected individuals ($\rho_I$) and aware individuals ($\rho_A$), respectively, under various coupling configurations. Dotted lines correspond to MMCA results, while scatter points indicate Monte Carlo (MC) simulation outcomes. In Figure~\ref{fig:couple}(a), under the condition where information exerts no effect on disease transmission ($\alpha=1$, $\Delta=1$), the infection levels $\rho_I$ are nearly identical for $\sigma = 1$ and $\sigma = 50$, as represented by the green and purple curves. However, the difference in $\rho_A$ is substantial in Figure~\ref{fig:couple}(b): when $\sigma = 50$, information reaches over 30\% of the population for $\lambda / \lambda_t > 1$, whereas for $\sigma = 1$, the information spread remains limited to 30\%, regardless of the infection rate.
When information does influence disease dynamics ($\alpha = 0.5$, $\Delta = 0.5$), a more complex interaction emerges. In Figure~\ref{fig:couple}(a), the infection level is significantly higher for $\sigma = 1$ than for $\sigma = 50$ once $\lambda / \lambda_t > 1$ (orange vs. blue curves), owing to the limited diffusion of awareness in the former case. Correspondingly, Figure~\ref{fig:couple}(b) shows that awareness reaches a much larger fraction of the population when $\sigma = 50$, leading to a stronger suppression of disease transmission. In summary, Figures~\ref{fig:couple}(a-b) reveal that stronger disease-to-information coupling (larger $\sigma$) markedly enhances awareness dissemination. In turn, increased public awareness (lower $\alpha$ and $\Delta$) feeds back into the system, reducing disease prevalence.

\begin{figure}
    \centering
    \includegraphics[width=0.7\linewidth]{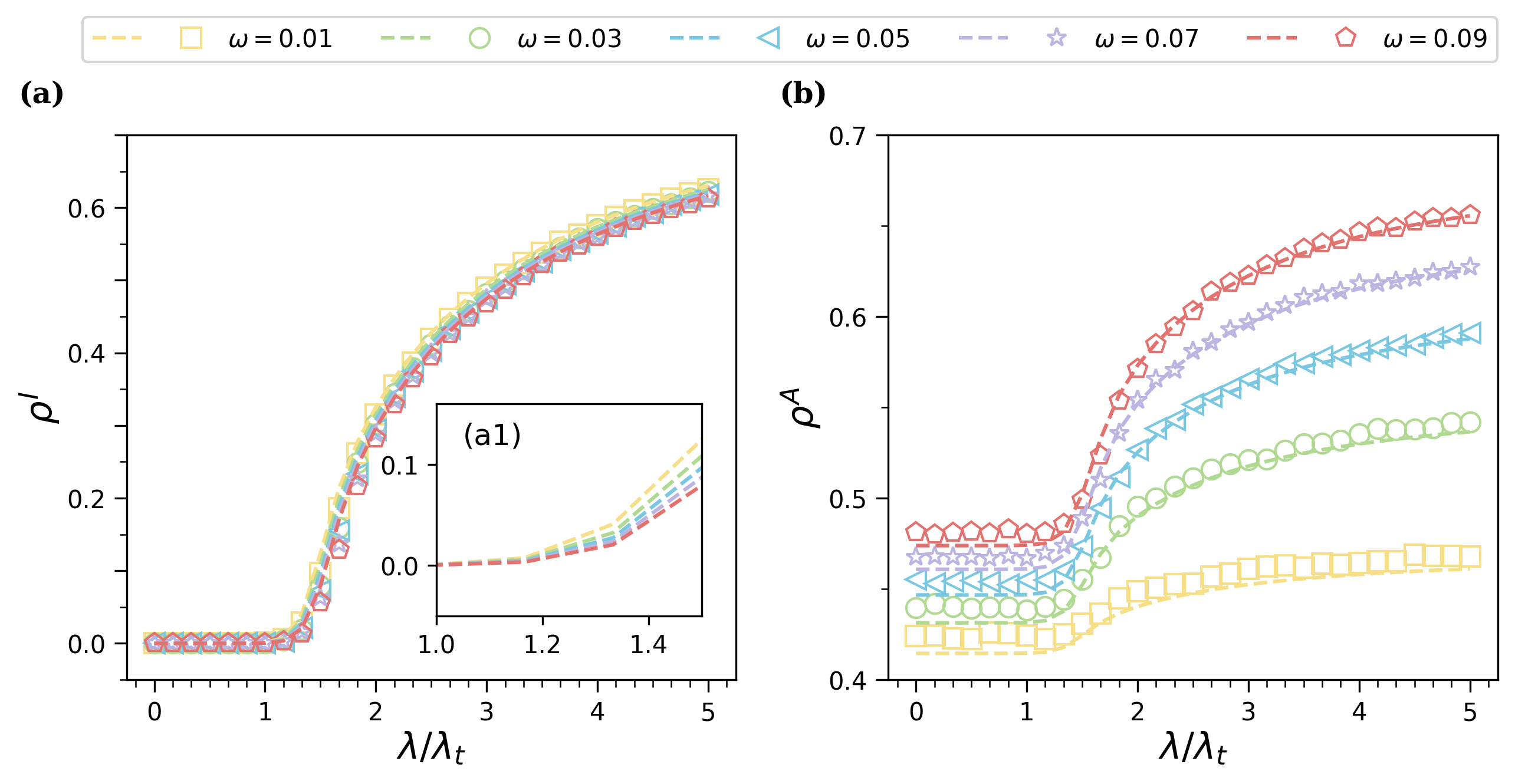}
    \caption{Impact of mass media on coupled spreading dynamics. (a) and (b) depict the steady-state proportions of A-state nodes ($\rho^{A}$) and I-state nodes ($\rho^{I}$), respectively, as functions of the normalized infection probability $\lambda / \lambda_{t}$ under different mass media transmission rates ($\omega$). Different colors correspond to varying values of $\omega$. Dotted lines indicate theoretical predictions derived from the MMCA approach, while scatter points represent the average results obtained from 200 independent MC simulation runs. (a1) shows a partial enlargement of $\rho^{I}$ using MMCA approach. The network consists of $N = 5000$ nodes, where the virtual information layer is generated by the BA model with $m=10$. The physical contact layer is modeled as a metapopulation network with $M=50$ subpopulations, each containing $Z=100$ nodes, with inter-subpopulation connections also generated using a BA model with $m=10$. Other parameters are set as $\sigma = 10$, $\beta_1 = 0.02$, $\beta_2 = 0.06$, $\beta_3 = 0.1$, $\gamma = 0.4$, $\alpha = 0.5$, $\mu = 0.3$, $g^{U} = 0.6$, and $\Delta = 0.5$.}
    \label{fig:media}
\end{figure}

We further examine the impact of mass media by varying the media broadcasting rate $\omega$, as shown in Figure \ref{fig:media}. Figures \ref{fig:media}(a) and \ref{fig:media}(b) illustrate the steady-state proportions of A-state and I-state nodes, respectively, under different values of $\omega$. As $\omega$ increases, Figure \ref{fig:media}(a) demonstrates a rise in the proportion of A-state nodes, indicating enhanced information awareness. Meanwhile, Figure \ref{fig:media}(b) shows a decline in I-state nodes accompanied by an increase in the epidemic threshold, suggesting that mass media dissemination effectively promotes protective behaviors, mitigates infection risks, and plays a critical role in disease control and prevention.

\begin{figure}
    \centering
    \includegraphics[width=0.7\linewidth]{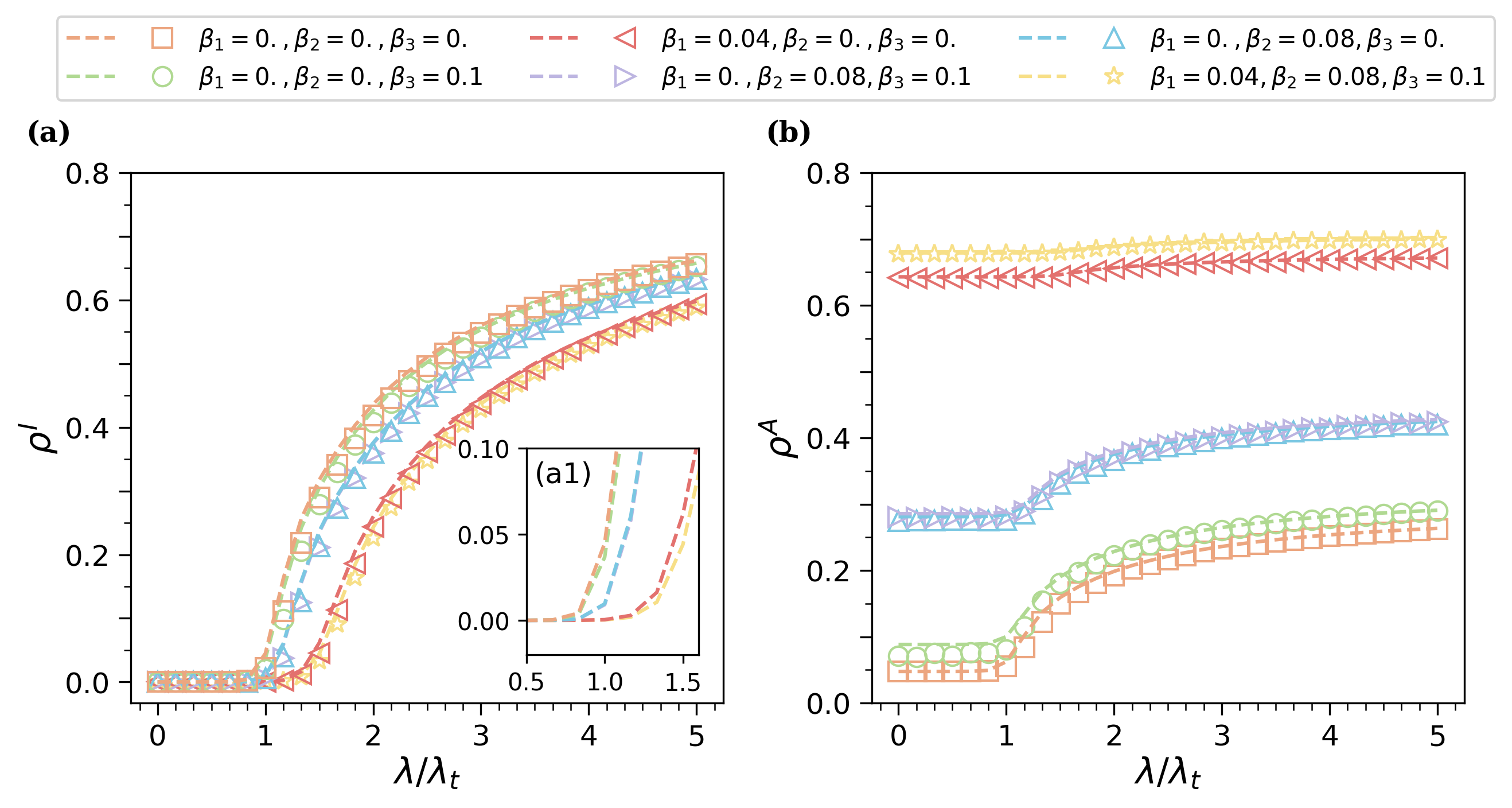}
    \caption{ Impact of simplex propagation rates on coupling dynamics. (a) and (b) show the steady-state proportions of A-state nodes ($\rho^{A}$) and I-state nodes ($\rho^{I}$), respectively, as functions of the normalized infection probability $\lambda / \lambda_{t}$ under different information dissemination parameters. Dotted lines represent theoretical predictions obtained using the MMCA approach, while scatter points correspond to the average results from 200 independent MC simulation runs. (a1) shows a partial enlargement of $\rho^{I}$ using MMCA approach. The network consists of $N = 5000$ nodes, where the virtual information layer follows a BA topology with $m=10$. The physical contact layer is structured as a metapopulation network with $M=50$ subpopulations, each containing $Z=100$ nodes, with inter-subpopulation connections also generated using a BA model with $m=10$. Other parameters are set as $\omega = 0.01$, $\sigma = 10$, $\gamma = 0.2$, $\alpha = 0.5$, $\mu = 0.3$, $g^{U} = 0.6$, and $\Delta = 0.5$.}
    \label{fig:ablation}
\end{figure}

The impact of higher-order structures on information diffusion is governed by two primary factors: the propagation rates associated with different orders of simplices and the number of these higher-order structures within the network. To systematically examine the effect of higher-order propagation rates, we fix the media transmission rate at $\omega = 0.01$ and design six experimental scenarios by varying the transmission rates of 1-simplices $(\beta_1)$, 2-simplices $(\beta_2)$, and 3-simplices $(\beta_3)$. The results are illustrated in Figure~\ref{fig:ablation}, where dotted lines correspond to the MMCA results and scatter points represent outcomes from MC simulations. Figures~\ref{fig:ablation}(a) and~\ref{fig:ablation}(b) show the steady-state fractions of infected individuals ($\rho_I$) and aware individuals ($\rho_A$), respectively, under different higher-order propagation configurations.
In Figure~\ref{fig:ablation}(a), relative to the baseline scenario involving only media influence (orange curve), the incorporation of higher-order structures, i.e., 1-simplices (red), 2-simplices (blue), 3-simplices (green), and their combinations (purple and yellow) effectively suppresses disease spread, reducing the final infection level by at least 5\%. Conversely, Figure~\ref{fig:ablation}(b) reveals that higher-order structures significantly boost the reach of information dissemination, enhancing $\rho_A$ by over 20\% in all cases compared to the media-only scenario. Interestingly, although 3-simplices contribute positively to both mitigating infection and promoting awareness, their marginal effect diminishes when 2-simplices are also present. This is evidenced by the overlap between the blue and purple curves. The underlying reason lies in the hierarchical nature of simplices, that is, 3-simplices inherently contain multiple 2-simplices. Thus, once the conditions for 3-simplex-based propagation are met, the corresponding 2-simplex pathways are also activated, reducing the exclusive influence of the 3-simplices.
Overall, the results demonstrate that, beyond the effect of media alone, the integration of higher-order interactions-across 1-, 2-, and 3-simplices-substantially enhances information diffusion while concurrently restraining epidemic outbreaks.

\begin{figure}
    \centering
    \includegraphics[width=1\linewidth]{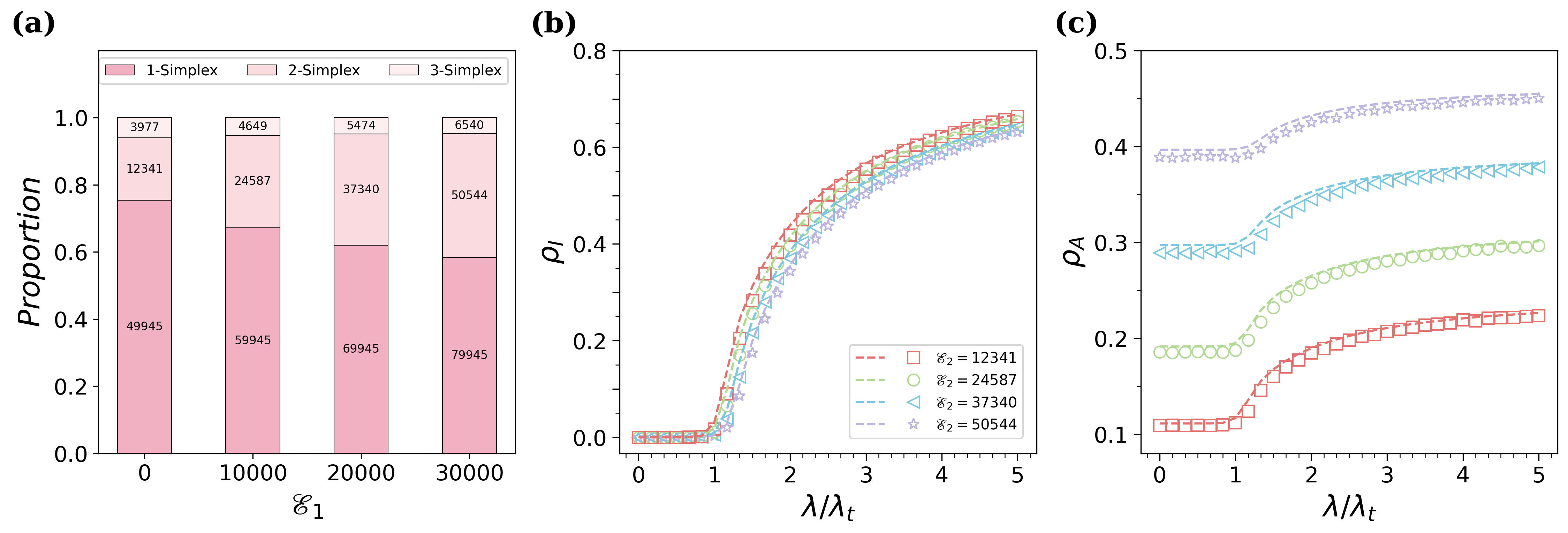}
    \caption{Impact of the number of 2-simplices on the coupled spreading dynamics.
(a) The proportions of 1-simplices, 2-simplices, and 3-simplices in the virtual information layer as a function of the number of added 1-simplices. (b) and (c) The steady-state fractions of infected ($\rho_I$) and aware ($\rho_A$) nodes, respectively, as functions of the normalized transmission rate $\lambda/\lambda_t$ under varying numbers of 2-simplices. Dotted lines represent theoretical predictions based on the MMCA method, while scatter points show averaged results over 200 MC simulations. The original network contains $N=5000$ nodes. The metapopulation structure in the physical contact layer follows a BA network with $m=10$, where each node represents a subpopulation of size $Z=100$. Model parameters are set as follows: $\omega = 0.01$, $\sigma = 10$, $\beta_1 = 0$, $\beta_2 = 0.08$, $\beta_3 = 0$, $\gamma = 0.4$, $\alpha = 0.5$, $\mu = 0.3$, $g^U = 0.6$, and $\Delta = 0.5$.}
    \label{fig:addsim}
\end{figure}

To investigate how the number of simplices in the virtual layer influences the coupled spreading dynamics, we systematically increase the number of 2-simplices and analyze their impact on the steady-state fractions of infected ($\rho_I$) and aware ($\rho_A$) nodes, as shown in Figure~\ref{fig:addsim}. Owing to the hierarchical nature of simplicial structures, increasing the number of 2-simplices requires first introducing additional 1-simplices (i.e., edges). Starting from a BA network with $N=5000$ nodes and $m=10$, we enhance the higher-order structures by randomly adding 1-simplices to the network. Let $\mathscr{E}_k$ denote the number of added $k$-simplices. The procedure for generating additional 2-simplices is outlined in Algorithm~\ref{algo:2}, where $D$ denotes the adjacency matrix of the original virtual-layer network. The algorithm selects a random pair of non-adjacent nodes that share a common neighbor and introduces an edge between them, thereby forming a 2-simplex. Figure~\ref{fig:addsim}(a) displays the evolution of the number and proportion of $k$-simplices with increasing 1-simplices. The baseline case of $\mathscr{E}_1=0$ corresponds to the unaltered BA network. As more 1-simplices are added, the numbers of both 2-simplices and 3-simplices naturally increase. To isolate the effect of 2-simplices and eliminate confounding factors, we fix the transmission rates associated with 1- and 3-simplices to zero (i.e., $\beta_1 = 0$ and $\beta_3 = 0$). Figures~\ref{fig:addsim}(b–c) illustrate how $\rho_I$ and $\rho_A$ vary with the number of 2-simplices. Dotted lines represent predictions from MMCA, while scatter plots show results from MC simulations. Our results indicate that higher-order structures in the virtual information layer exert a strong influence on the coupled spreading dynamics. As the number of 2-simplices increases, the prevalence of infection decreases (Figure~\ref{fig:addsim}(b)), while the level of awareness rises (Figure~\ref{fig:addsim}(c)). This suggests that enhancing group-level interactions within information networks can improve public awareness and effectively mitigate disease transmission. In conclusion, during infectious disease outbreaks, fostering group-based communication and collaboration can play a vital role in elevating preventive awareness and curbing epidemic spread.

\begin{algorithm}
\caption{ Random Addition of 1-simplices to Increase 2-simplices}
\label{algo:2}
\KwIn{Adjacency matrix $D$, The number of added 1-simplices $\mathscr{E}_{1}$}
\KwOut{Updated adjacency matrix $D$}

The number of 1-simplices has been increased $n \gets 0$\;

\While{$n < E_{A}$}{
    Randomly select two distinct nodes $i$ and $j$\;
    
    \If{$D[i,j] = 0$}{
        Find common neighbors of $i$ and $j$\;
        
        \If{common neighbors exist}{
            $D[i,j] = D[j,i] = 1$\;
            $n=n+1$\;
        }
    }
}
\Return{Updated adjacency matrix $D$}
\end{algorithm}

\begin{figure}
    \centering
    \includegraphics[width=0.7\linewidth]{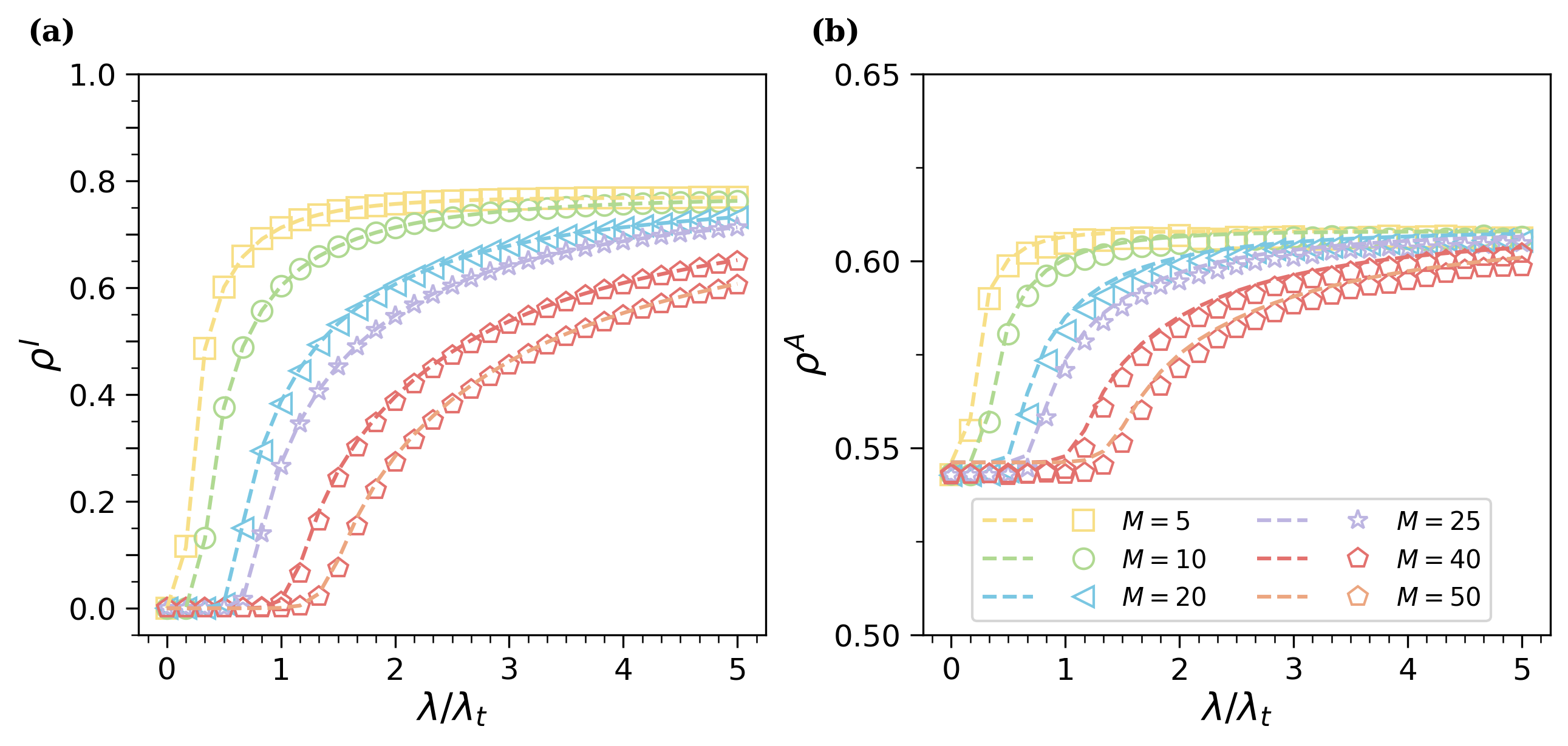}
    \caption{Effect of subpopulation number on coupled spreading dynamics.
(a) and (b) show the steady-state fractions of aware (A-state) and infected (I-state) nodes, respectively, as functions of the normalized transmission rate $\lambda/\lambda_t$ for different numbers of subpopulations in the physical contact layer. Dotted lines indicate theoretical predictions based on the MMCA method, while scatter points represent the average results from 200 MC simulations. The virtual information layer is constructed as a BA network with $m=10$ and $N=5000$ nodes. Other parameters are set as follows: $\omega = 0.01$, $\sigma = 10$, $\beta_1 = 0.04$, $\beta_2 = 0.08$, $\beta_3 = 0.1$, $\gamma = 0.4$, $\alpha = 0.5$, $\mu = 0.3$, $g^U = 0.6$, and $\Delta = 0.5$.}
    \label{fig:changenum}
\end{figure}

We further explore how the metapopulation structure influences the coupled dynamics by examining the effects of the number of subpopulations, the overall network structure, and the heterogeneity within the metapopulation network. For metapopulation networks with a BA topology comprising $N=5000$ nodes, we fix the parameter $m=3$ to maintain a consistent link density. By varying the number of subpopulations $M$ while ensuring equal subpopulation sizes ($Z_k = N/M$), we analyze how this structural adjustment affects disease propagation, as illustrated in Figure~\ref{fig:changenum}. Figures~\ref{fig:changenum}(a) and \ref{fig:changenum}(b) show the steady-state fractions of infected individuals ($\rho^I$) and aware individuals ($\rho^A$), respectively, under different values of $M$. Dotted lines represent the results obtained from the MMCA framework, while scatter points correspond to averages over MC simulations. Figure~\ref{fig:changenum}(a) reveals that a smaller number of subpopulations leads to higher levels of infection and allows the disease to spread more easily, even at lower transmission rates $\lambda$ (as seen in the yellow curve). This outcome arises because fewer subpopulations result in more individuals within each subpopulation, thereby intensifying local transmission. In turn, this elevated disease prevalence promotes broader information dissemination, as shown in Figure~\ref{fig:changenum}(b). Overall, Figures~\ref{fig:changenum}(a–b) demonstrate that reducing (increasing) the number of subpopulations $M$ amplifies (mitigates) the spread of both the disease and associated awareness. These findings suggest that limiting large gatherings and maintaining decentralized population structures may be effective strategies for controlling outbreaks.


\begin{figure}
    \centering
    \includegraphics[width=1\linewidth]{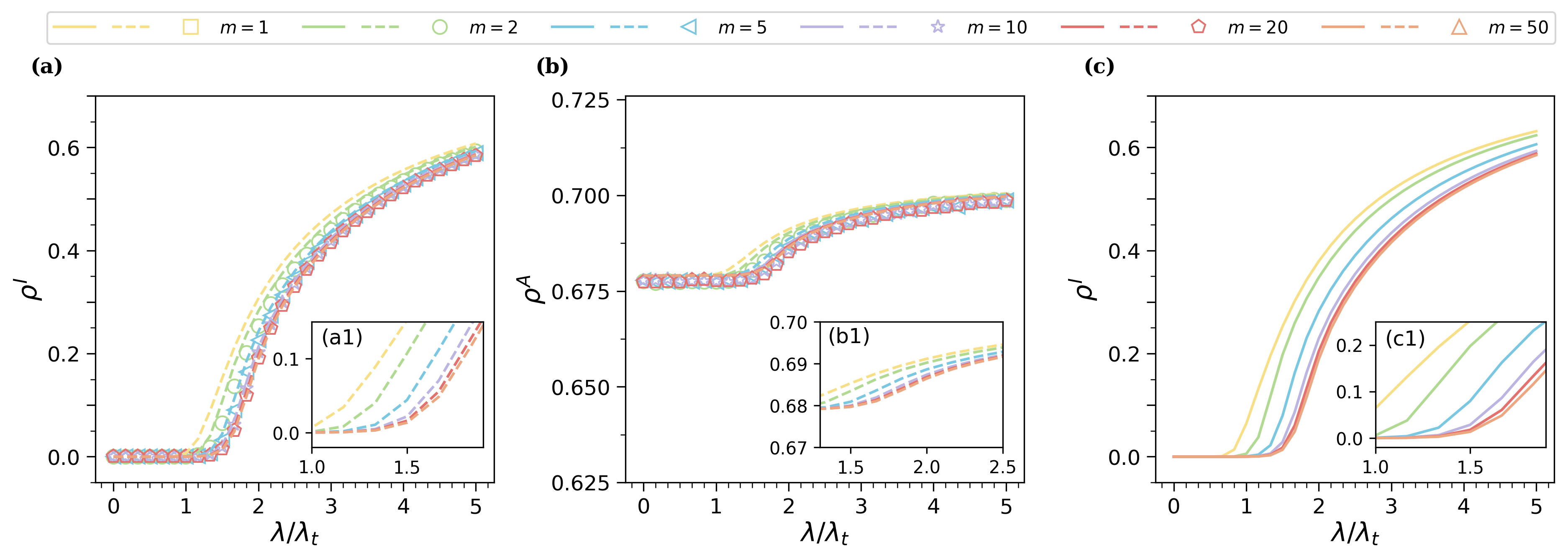}
    \caption{Impact of metapopulation network connectivity on spreading dynamics. Panels (a) and (b) depict the steady-state proportions of infected ($\rho^{I}$) and aware ($\rho^{A}$) nodes, respectively, as functions of the normalized transmission rate $\lambda/\lambda_t$ across metapopulation networks with varying levels of connectivity (i.e., different values of $m$). (a1) and (b1) provide magnified views of the corresponding curves obtained from the MMCA. In all subplots, dotted lines indicate theoretical predictions derived from MMCA, while scatter points represent averaged results from 200 MC simulations. (c) shows the steady-state infection levels ($\rho^I$) in the absence of awareness-based movement, also plotted as functions of $\lambda/\lambda_t$ for different values of $m$. (c1) presents a zoomed-in view of $\rho^I$ under this setting, based on MMCA predictions. In these simulations, awareness is neutralized by setting $R^A = R^U$ and $\Delta = 1$. The virtual information layer is modeled as a BA network with $m = 10$ and $N = 5000$ nodes. The remaining model parameters are set as: $\omega = 0.01$, $\sigma = 10$, $\beta_1 = 0.04$, $\beta_2 = 0.08$, $\beta_3 = 0.1$, $\gamma = 0.4$, $\alpha = 0.5$, $\mu = 0.3$, and $g^U = 0.6$.}
    \label{fig:changem}
\end{figure}

\begin{figure}
    \centering
    \includegraphics[width=1\linewidth]{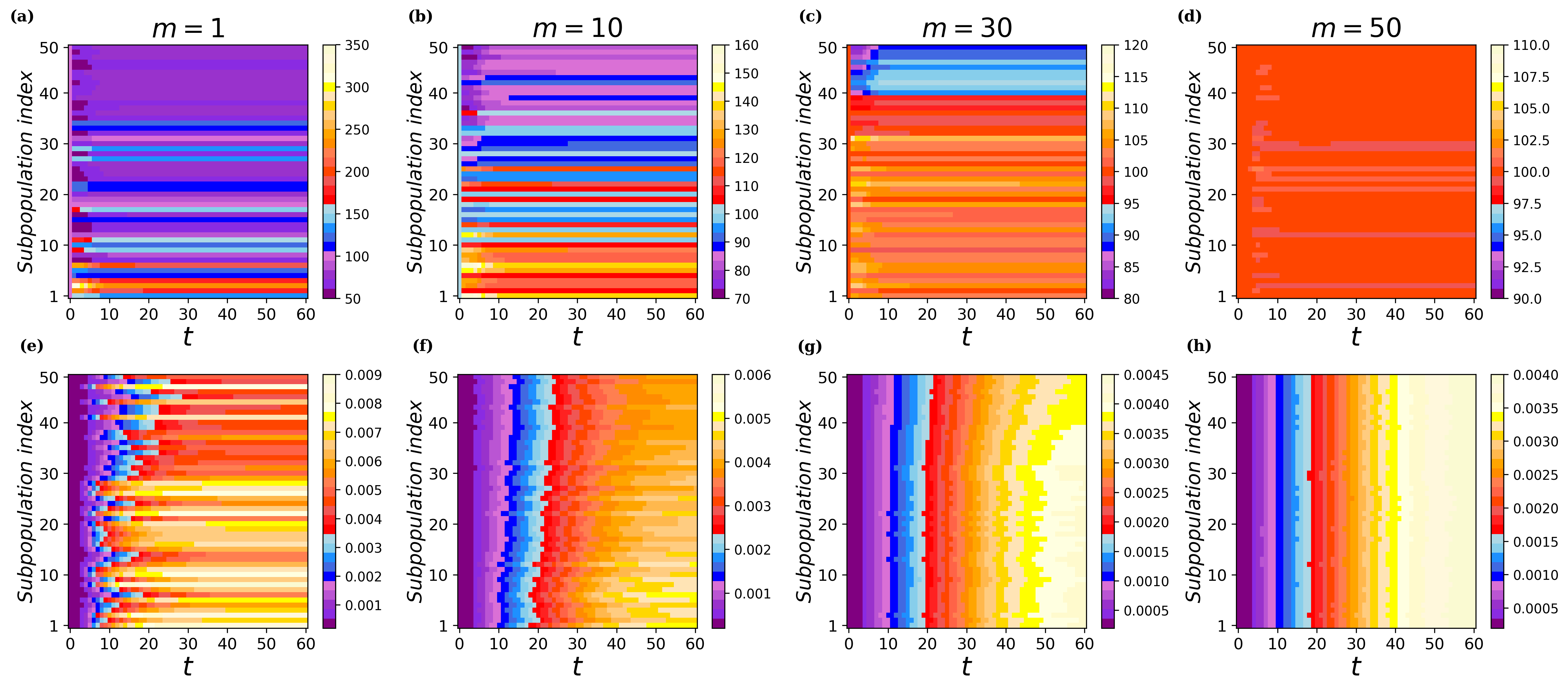}
    \caption{Heatmaps illustrating the impact of metapopulation network connectivity on migration patterns and epidemic dynamics. (a–d) display the temporal evolution of the number of nodes in each subpopulation after the movement phase, for BA-structured metapopulation networks with $m = 1$, $m = 10$, $m = 30$, and $m = 50$, respectively. (e–h) show the corresponding steady-state proportions of infected individuals ($\rho^I$) in each subpopulation after the return phase, calculated using MMCA. The results reflect how increasing connectivity leads to more uniform node distributions and a reduction in local outbreak intensities. The model parameters are set as follows: $\omega = 0.01$, $\sigma = 10$, $\beta_1 = 0.04$, $\beta_2 = 0.08$, $\beta_3 = 0.1$, $\gamma = 0.2$, $\lambda = 0.06$, $\alpha = 0.5$, $\mu = 0.3$, $g^U = 0.6$, and $\Delta = 0.5$.}
    \label{fig:changeedgenum}
\end{figure}

To investigate the impact of subpopulation connectivity on coupled spreading dynamics, we construct BA metapopulation networks consisting of $N = 5000$ nodes distributed across $M = 50$ subpopulations. By tuning the parameter $m$, which determines the number of connections a new node establishes upon entering the network, we explore how variations in inter-subpopulation connectivity influence transmission patterns, as shown in Figure~\ref{fig:changem}. Specifically, Figures~\ref{fig:changem}(a) and \ref{fig:changem}(b) illustrate the steady-state fractions of infected individuals ($\rho^I$) and aware individuals ($\rho^A$), respectively, across different values of $m$. As shown in Figure~\ref{fig:changem}(a), increasing $m$ results in a reduction of $\rho^I$, suggesting that enhanced connectivity suppresses infection spread. Similarly, Figure~\ref{fig:changem}(b) reveals a decrease in $\rho^A$, which can be attributed to the lowered infection prevalence, thereby reducing the activation of awareness. To confirm that this effect is not primarily driven by the awareness mechanism itself, we perform a control experiment by disabling awareness (i.e., setting $R^A=R^U$ and $\Delta=1$). As shown in Figure~\ref{fig:changem}(c), the downward trend in $\rho^I$ with increasing $m$ persists even in the absence of awareness, indicating that this phenomenon stems from structural and dynamic properties of the network rather than awareness alone. To further clarify this counterintuitive finding that increased connectivity can lead to decreased infection prevalence, we analyze how inter-subpopulation connectivity ($m$) affects node mobility and the resulting epidemic dynamics, as shown in Figure~\ref{fig:changeedgenum}. Figures~\ref{fig:changeedgenum}(a–d) display the number of nodes $Z_k$ in each subpopulation at the end of the mobility phase, for networks with $m = 1$, $10$, $30$, and $50$, respectively. Figures~\ref{fig:changeedgenum}(e–h) show the corresponding infection levels $\rho^I$ within each subpopulation after the return phase. As illustrated in Figures~\ref{fig:changeedgenum}(a–d), higher values of $m$ lead to a more uniform distribution of nodes across subpopulations. In the case of $m = 1$, mobility results in pronounced heterogeneity, with subpopulation 3 consistently attracting more than 200 nodes (Figure~\ref{fig:changeedgenum}(a)). In contrast, when $m = 50$, indicating nearly complete interconnectivity, subpopulations exhibit uniform sizes, each containing approximately $N/M = 100$ nodes (Figure~\ref{fig:changeedgenum}(d)). Figures~\ref{fig:changeedgenum}(e–h) further demonstrate that as $m$ increases, the infection fractions $\rho^I$ across subpopulations tend to converge, and the overall steady-state infection level (at $t = 60$) declines. For low values of $m$, the network’s heterogeneity allows certain subpopulations, such as subpopulation 3 in the $m = 1$ case, to accumulate disproportionately large populations during mobility, enabling localized outbreaks and amplifying disease prevalence. In contrast, a higher $m$ results in evenly distributed populations, reducing the risk of concentrated outbreaks and thus suppressing the overall spread. In conclusion, Figures~\ref{fig:changem} and \ref{fig:changeedgenum} collectively show that increasing inter-subpopulation connectivity in a connected metapopulation network leads to more balanced node distributions after migration, which in turn mitigates epidemic prevalence. Moreover, although the awareness mechanism can significantly alter transmission dynamics by influencing mobility patterns, the observed decline in infection levels with increasing $m$ remains valid even in its absence, underscoring the critical role of structural connectivity in disease dynamics.

\begin{figure}
    \centering
    \includegraphics[width=1\linewidth]{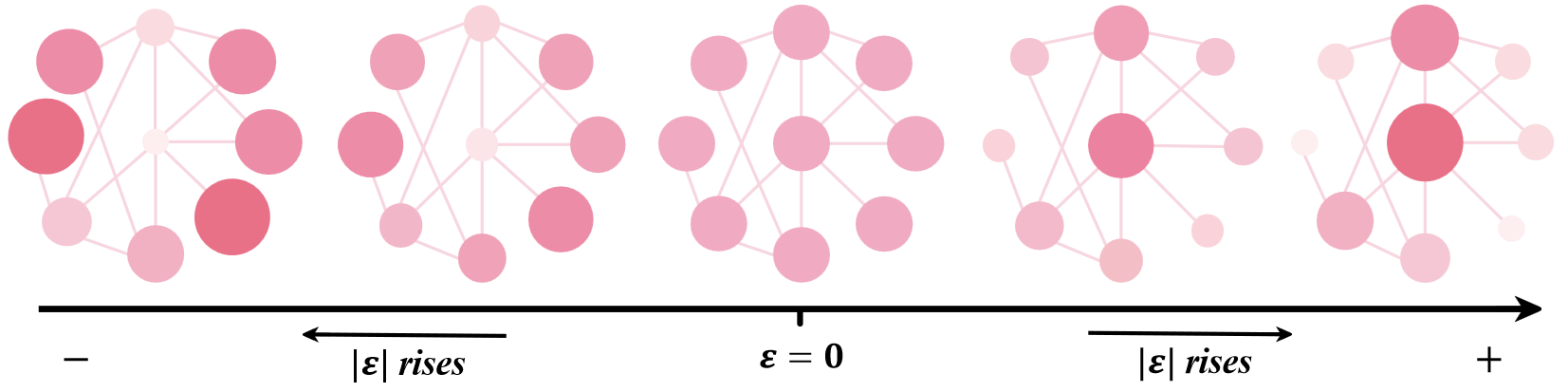}
    \caption{Illustrative diagram of metapopulation network structure under varying $\varepsilon$.  Each circle denotes a subpopulation, where a larger size and darker red color indicate a greater number of nodes. When $\varepsilon=0$, nodes are evenly distributed, and all subpopulations contain the same number of nodes. For $\varepsilon>0$, subpopulations with higher degrees (i.e., more inter-subpopulation connections) host more nodes, with the disparity growing as $|\varepsilon|$  increases. In contrast, when $\varepsilon<0 $, subpopulations with fewer connections are assigned more nodes, and this imbalance also becomes more pronounced as $|\varepsilon|$ increases.}
    \label{fig:Heterogeneity diagram}
\end{figure}

\begin{figure}
    \centering
    \includegraphics[width=0.7\linewidth]{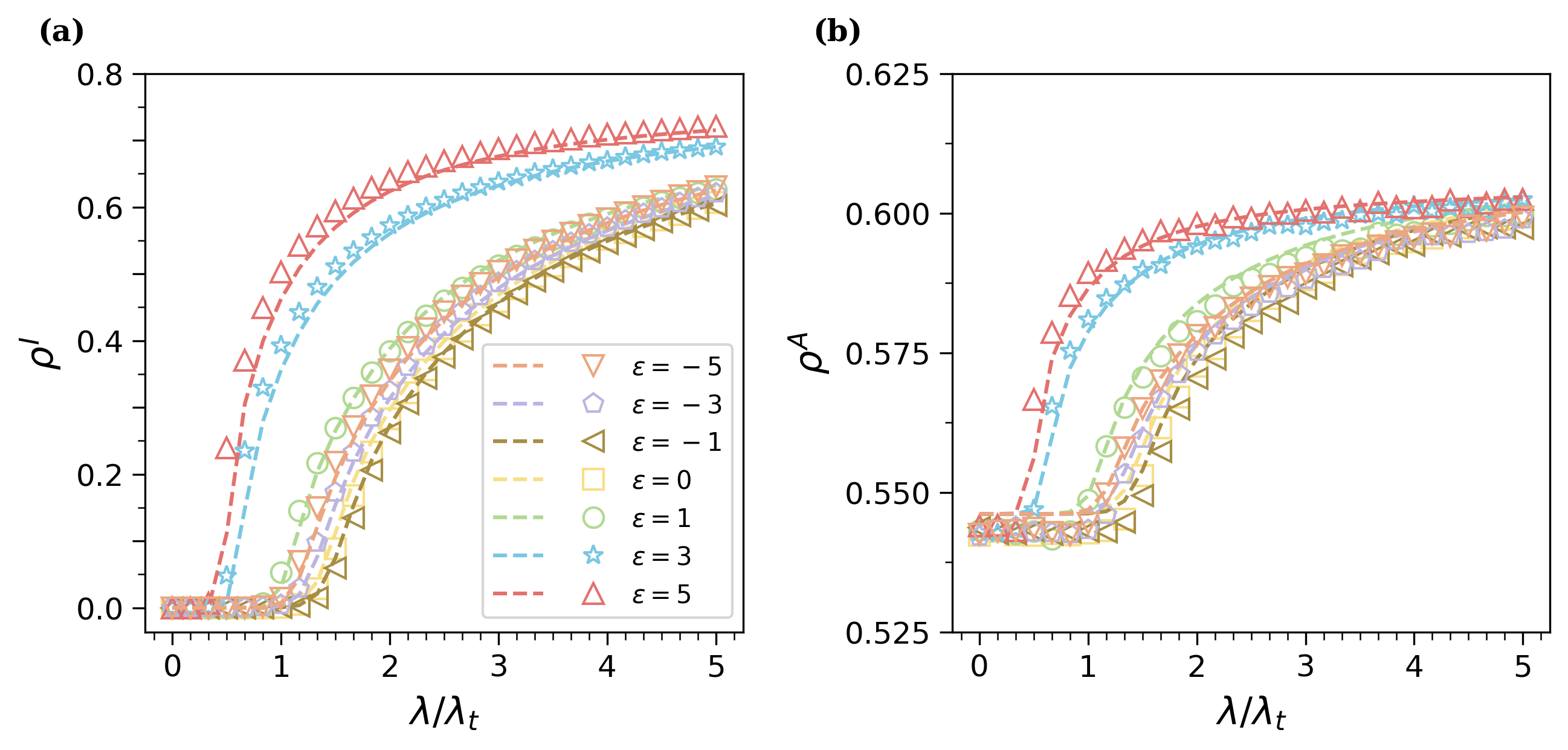}
    \caption{Steady-state fractions of infected (a) and aware (b) nodes versus normalized transmission rate for various subpopulation heterogeneity levels ($\varepsilon$). Theoretical curves (MMCA) are shown as dotted lines; simulation averages (200 MC runs) as scatter points. The information layer is a BA network ($N=5000$, $m=10$); the contact layer is a BA-based metapopulation network with $M=50$ subpopulations. Parameters: $\omega = 0.01$, $\sigma = 10$, $\beta_1 = 0.04$, $\beta_2 = 0.08$, $\beta_3 = 0.1$, $\gamma = 0.4$, $\alpha = 0.5$, $\mu = 0.3$, $g^U = 0.6$, $\Delta = 0.5$.}
    \label{fig:heterogeneity}
\end{figure}

The heterogeneity of subpopulations refers to the variation in the number of nodes across different subpopulations. To investigate the impact of such heterogeneity on coupling dynamics, we modify the structural configuration of the metapopulation. Let 
$D_k$ denote the degree of subpopulation $k$, and $Z_k$ the number of nodes within subpopulation $k$. We define $Z_k$ as follows:

\begin{equation}
Z_{k}=N\cdot (D_{k}^\varepsilon /\sum_{l=1}^{M}D_{l}^\varepsilon ),
\label{formula:21}
\end{equation}
where the parameter $\varepsilon$ governs the correlation between  $D_{k}$ and $Z_{k}$.
As shown in Figure \ref{fig:Heterogeneity diagram}, when $\varepsilon=0$, there is no correlation, and all subpopulations have an equal number of nodes, i.e., $Z_{k}= N/M$. When $\varepsilon>0$, subpopulations with higher degrees (i.e., larger $D_k$) tend to contain more nodes (i.e., larger $Z_k$), with this effect becoming more pronounced as $|\varepsilon|$ increases. In contrast, when $\varepsilon<0 $, highly connected subpopulations host fewer nodes, and the disparity grows with increasing $|\varepsilon|$. Figures~\ref{fig:heterogeneity}(a) and \ref{fig:heterogeneity}(b) display the steady-state fractions of infected nodes ($\rho^{I}$) and aware nodes ($\rho^{A}$), respectively, across different values of $\varepsilon$. In both figures, the dotted lines represent results obtained from the MMCA approach, while the scatter points correspond to MC simulation results. When $\varepsilon > 0$, $\rho^{I}$ increases significantly with the absolute value of $\varepsilon$, as shown in Figure~\ref{fig:heterogeneity}(a). Notably, when $\varepsilon = 3$ (blue curve), the infected fraction $\rho^{I}$ at $\lambda / \lambda_{t} = 1$ is approximately 25\% higher compared to the case of $\varepsilon = 1$ (green curve). Similarly, Figure~\ref{fig:heterogeneity}(b) shows that $\rho^{A}$ also rises with increasing $|\varepsilon|$ when $\varepsilon > 0$, which is attributed to the higher overall infection prevalence. In the case of $\varepsilon < 0$, both $\rho^{I}$ and $\rho^{A}$ also increase as $|\varepsilon|$ grows, as seen in Figures~\ref{fig:heterogeneity}(a) and \ref{fig:heterogeneity}(b). However, the overall infected proportion remains lower compared to scenarios with positive $\varepsilon$ values (e.g., $\varepsilon = 1, 3, 5$). This behavior arises because, when $\varepsilon < 0$, subpopulations with fewer interconnections tend to have larger node counts. Although higher local densities promote within-group transmission, the limited connectivity suppresses inter-subpopulation disease spread. Interestingly, when $\varepsilon = -1$ (brown curve), the infected fraction $\rho^{I}$ slightly falls short of that of the homogeneous case $\varepsilon = 0$ (yellow curve), suggesting that modestly directing individuals toward regions with restricted mobility could help mitigate epidemic outbreaks. Overall, Figure~\ref{fig:heterogeneity} highlights that the heterogeneity parameter $\varepsilon$ significantly impacts the coupled spreading dynamics by altering the node distribution across subpopulations. The disease reaches its minimal transmission extent when $\varepsilon = -1$. These findings imply that avoiding densely connected areas and limiting movement between regions are effective strategies for epidemic control during infectious disease outbreaks.

\begin{figure}
    \centering
    \includegraphics[width=1\linewidth]{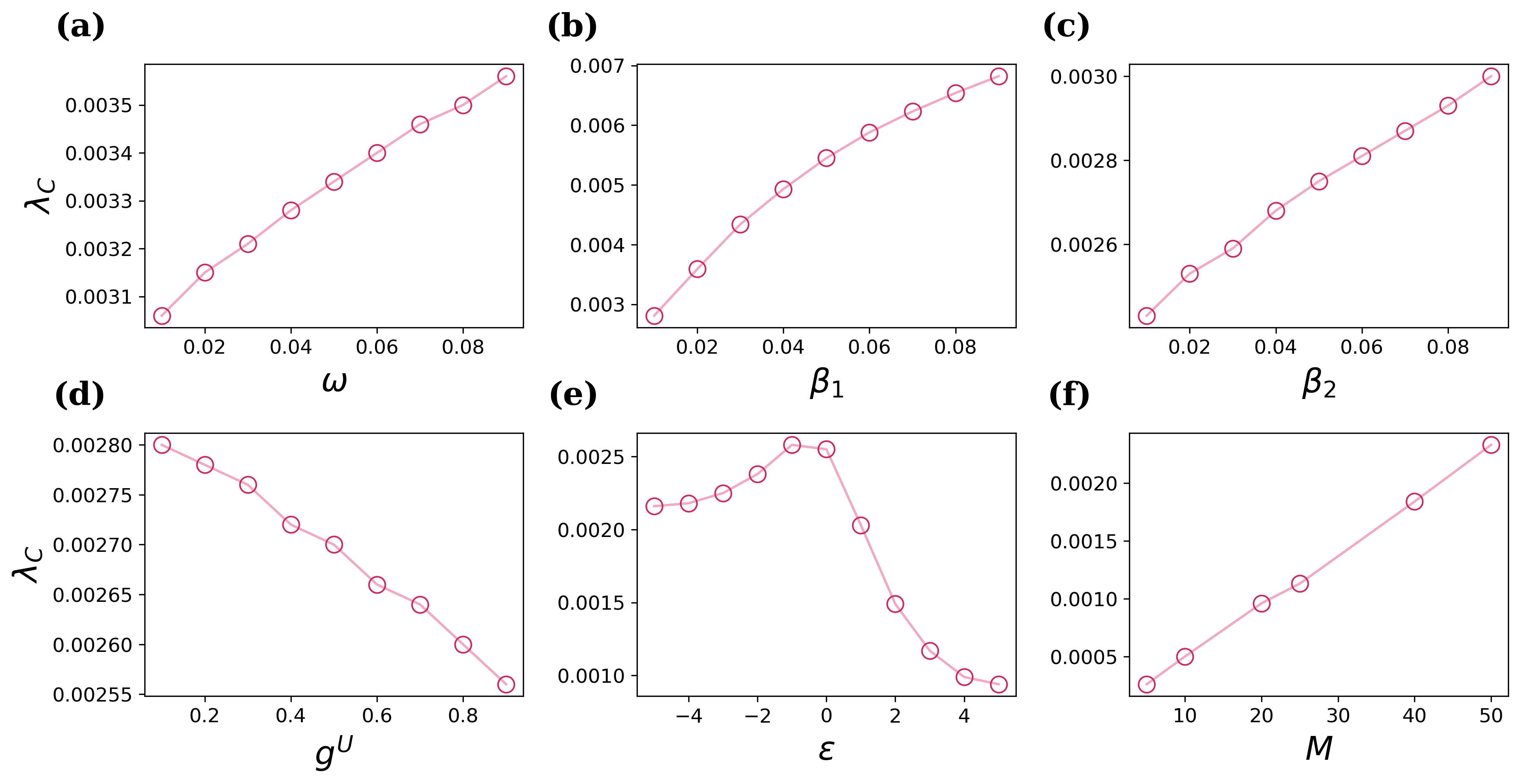}
    \caption{Influence of model parameters on the epidemic threshold $\lambda_c$. (a-c) illustrate the effects of structural and dynamical parameters in the information layer, specifically: the media broadcasting rate $\omega$, and the transmission rates over 1-simplices ($\beta_1$) and 2-simplices ($\beta_2$), respectively. (d-f) present the influence of key factors in the contact layer, including the node migration rate $g^U$, the subpopulation heterogeneity parameter $\varepsilon$, and the number of subpopulations $M$, respectively. Unless otherwise specified, the parameters are set as: $\sigma = 10$, $\beta_3 = 0.1$, $\gamma = 0.4$, $\alpha = 0.5$, $\mu = 0.3$, and $\Delta = 0.5$.}
    \label{fig:threshold}
\end{figure}

To intuitively explore how model parameters influence the epidemic threshold, we separately analyze the structural impacts of the information layer and the contact layer, as illustrated in Figure~\ref{fig:threshold}. Figures~\ref{fig:threshold}(a–c) examine the effects of the information layer. As shown in Figure~\ref{fig:threshold}(a), the epidemic threshold $\lambda_c$ increases with the media broadcasting rate $\omega$, underscoring the critical role of media-driven awareness in suppressing epidemic outbreaks. Similarly, Figures~\ref{fig:threshold}(b) and \ref{fig:threshold}(c) reveal that $\lambda_c$ rises with increasing transmission rates over 1-simplices ($\beta_1$) and 2-simplices ($\beta_2$), respectively, highlighting that higher-order information transmission significantly contributes to disease containment. Figures~\ref{fig:threshold}(d–f) focus on the contact layer. In Figure~\ref{fig:threshold}(d), the epidemic threshold $\lambda_c$ decreases as the node migration rate $g^U$ increases, suggesting that greater population mobility facilitates disease propagation and lowers the outbreak threshold. Figure~\ref{fig:threshold}(e) demonstrates a nonlinear relationship between subpopulation heterogeneity (parameterized by $\varepsilon$) and the epidemic threshold, i.e., $\lambda_c$ increases with $\varepsilon$ when $\varepsilon < 0$, but decreases when $\varepsilon > 0$. This implies that both extremely homogeneous and extremely heterogeneous population structures can have distinct effects on epidemic control. Figure~\ref{fig:threshold}(f) shows that increasing the number of subpopulations $M$ leads to a higher epidemic threshold, indicating improved epidemic resistance in more fragmented population structures. In summary, Figure~\ref{fig:threshold} reveals that the epidemic threshold $\lambda_c$ is positively associated with media broadcasting strength, higher-order information interactions, and subpopulation count, while being negatively correlated with migration intensity. Notably, the impact of subpopulation heterogeneity on the threshold is nonlinear, emphasizing the complexity of structural influences in metapopulation epidemic dynamics.

\section{Conclusion}
The COVID-19 pandemic has highlighted the complexity of infectious disease transmission, demonstrating that it is a multifactorial process influenced by media effects, information diffusion, and population mobility. While previous studies have explored these factors individually, few have integrated them into a unified framework that reflects the interplay of real-world dynamics. To bridge this gap, we proposed a novel dynamic model of infectious disease–information coevolution that incorporates metapopulation structures and mass media influence.

Our model is constructed on a two-layer network. The upper layer represents a virtual information space modeled as a higher-order simplicial complex network, where each node is also linked to a media node to capture mass media effects. The lower layer represents physical disease transmission across a metapopulation network based on a reaction–diffusion process. We derived the MMCA equations to describe the coupled spreading dynamics and validated them against extensive MC simulations. The results reveal strong coupling between information and disease spreading, i.e., information diffusion can effectively suppress the spread of disease, while the rise in infection rates boosts public awareness, thereby amplifying information dissemination. Media influence and higher-order simplices play a critical role in shaping the diffusion of information. The contribution of a simplex to the co-spreading process tends to diminish as its order increases. Moreover, our results reveal that increasing the number of edges in a BA-structured connected metapopulation network leads to a reduction in overall disease prevalence. Additionally, both population mobility and the internal heterogeneity of subpopulations are shown to exert substantial influence on epidemic dynamics. Interestingly, our findings suggest that, under certain conditions, moderately directing individuals to migrate toward less connected or more isolated regions can effectively help suppress the spread of infection.


Despite these insights, our study has several limitations that suggest avenues for future research. First, the model assumes that all disseminated information is accurate, thereby neglecting the potential impact of misinformation or conflicting messages\cite{guo2020future}. Second, the disease spreading dynamics are modeled using a simple SIS framework, which may not capture the full complexity of real-world epidemic, such as diseases with multiple stages and varying infectiousness (e.g., HIV/AIDS). Third, subpopulations are modeled as fully connected graphs, overlooking the heterogeneity inherent in real social networks. In addition, node connections in the virtual information layer are assumed to be independent of their physical subpopulation affiliations, potentially missing cross-layer correlations found in real systems. Finally, both the physical and virtual network topologies are assumed to be static, whereas incorporating temporal network dynamics remains an important direction for future exploration\cite{zhan2025measuring}.
\section*{Acknowledgment}
This work was supported by the China Postdoctoral Science Foundation (2024M762809), the National Natural Science Foundation of China (Grant No. 62473123), the Scientific Research Foundation for Scholars of HZNU (2021QDL030), and the Scientific Research Fund of Zhejiang Provincial Education Department (Y202454842).
\bibliography{ref.bib}

\begin{thebibliography}{10}

\bibitem{keeling2005networks}
Matt~J Keeling and Ken~TD Eames.
\newblock Networks and epidemic models.
\newblock {\em Journal of the royal society interface}, 2(4):295--307, 2005.

\bibitem{lu2021stability}
Xing L{\"u}, Hong-wen Hui, Fei-fei Liu, and Ya-li Bai.
\newblock Stability and optimal control strategies for a novel epidemic model
  of covid-19.
\newblock {\em Nonlinear Dynamics}, 106(2):1491--1507, 2021.

\bibitem{wang2017unification}
Wei Wang, Ming Tang, H~Eugene Stanley, and Lidia~A Braunstein.
\newblock Unification of theoretical approaches for epidemic spreading on
  complex networks.
\newblock {\em Reports on progress in physics}, 80(3):036603, 2017.

\bibitem{li2021dynamics}
Hui-Jia Li, Wenzhe Xu, Shenpeng Song, Wen-Xuan Wang, and Matja{\v{z}} Perc.
\newblock The dynamics of epidemic spreading on signed networks.
\newblock {\em Chaos, Solitons \& Fractals}, 151:111294, 2021.

\bibitem{basnarkov2021seair}
Lasko Basnarkov.
\newblock Seair epidemic spreading model of covid-19.
\newblock {\em Chaos, Solitons \& Fractals}, 142:110394, 2021.

\bibitem{boccaletti2020modeling}
Stefano Boccaletti, William Ditto, Gabriel Mindlin, and Abdon Atangana.
\newblock Modeling and forecasting of epidemic spreading: The case of covid-19
  and beyond.
\newblock {\em Chaos, solitons, and fractals}, 135:109794, 2020.

\bibitem{morris2021optimal}
Dylan~H Morris, Fernando~W Rossine, Joshua~B Plotkin, and Simon~A Levin.
\newblock Optimal, near-optimal, and robust epidemic control.
\newblock {\em Communications Physics}, 4(1):78, 2021.

\bibitem{zhan2018coupling}
Xiu-Xiu Zhan, Chuang Liu, Ge~Zhou, Zi-Ke Zhang, Gui-Quan Sun, Jonathan~JH Zhu,
  and Zhen Jin.
\newblock Coupling dynamics of epidemic spreading and information diffusion on
  complex networks.
\newblock {\em Applied Mathematics and Computation}, 332:437--448, 2018.

\bibitem{anwar2020role}
Ayesha Anwar, Meryem Malik, Vaneeza Raees, and Anjum Anwar.
\newblock Role of mass media and public health communications in the covid-19
  pandemic.
\newblock {\em Cureus}, 12(9), 2020.

\bibitem{zhang2023study}
Jing Zhang, Xiaoli Wang, and Shuqin Chen.
\newblock Study on the interaction between information dissemination and
  infectious disease dissemination under government prevention and management.
\newblock {\em Chaos, Solitons \& Fractals}, 173:113601, 2023.

\bibitem{wu2024influence}
Bingjie Wu et~al.
\newblock The influence of different government policies on the co-evolution of
  information dissemination, vaccination behavior and disease transmission in
  multilayer networks.
\newblock {\em Chaos, Solitons \& Fractals}, 180:114522, 2024.

\bibitem{kabir2019analysis}
KM~Ariful Kabir, Kazuki Kuga, and Jun Tanimoto.
\newblock Analysis of sir epidemic model with information spreading of
  awareness.
\newblock {\em Chaos, Solitons \& Fractals}, 119:118--125, 2019.

\bibitem{xu2015epidemic}
Qichao Xu, Zhou Su, Kuan Zhang, Pinyi Ren, and Xuemin~Sherman Shen.
\newblock Epidemic information dissemination in mobile social networks with
  opportunistic links.
\newblock {\em IEEE Transactions on Emerging Topics in Computing},
  3(3):399--409, 2015.

\bibitem{funk2009spread}
Sebastian Funk, Erez Gilad, Chris Watkins, and Vincent~AA Jansen.
\newblock The spread of awareness and its impact on epidemic outbreaks.
\newblock {\em Proceedings of the National Academy of Sciences},
  106(16):6872--6877, 2009.

\bibitem{han2024impact}
Dun Han and Xin Wang.
\newblock Impact of positive and negative information on epidemic spread in a
  three-layer network.
\newblock {\em Chaos, Solitons \& Fractals}, 186:115264, 2024.

\bibitem{huang2021modeling}
He~Huang, Yahong Chen, and Yefeng Ma.
\newblock Modeling the competitive diffusions of rumor and knowledge and the
  impacts on epidemic spreading.
\newblock {\em Applied mathematics and computation}, 388:125536, 2021.

\bibitem{kabir2019effect}
KM~Ariful Kabir, Kazuki Kuga, and Jun Tanimoto.
\newblock Effect of information spreading to suppress the disease contagion on
  the epidemic vaccination game.
\newblock {\em Chaos, Solitons \& Fractals}, 119:180--187, 2019.

\bibitem{granell2013dynamical}
Clara Granell, Sergio G{\'o}mez, and Alex Arenas.
\newblock Dynamical interplay between awareness and epidemic spreading in
  multiplex networks.
\newblock {\em Physical review letters}, 111(12):128701, 2013.

\bibitem{hong2022co}
Xiao Hong, Yuexing Han, Gouhei Tanaka, and Bing Wang.
\newblock Co-evolution dynamics of epidemic and information under dynamical
  multi-source information and behavioral responses.
\newblock {\em Knowledge-Based Systems}, 252:109413, 2022.

\bibitem{xia2024dynamic}
Yang Xia, Haijun Jiang, Shuzhen Yu, and Zhiyong Yu.
\newblock The dynamic analysis of the rumor spreading and behavior diffusion
  model with higher-order interactions.
\newblock {\em Communications in Nonlinear Science and Numerical Simulation},
  138:108186, 2024.

\bibitem{pei2015exploring}
Sen Pei, Lev Muchnik, Shaoting Tang, Zhiming Zheng, and Hern{\'a}n~A Makse.
\newblock Exploring the complex pattern of information spreading in online blog
  communities.
\newblock {\em PloS one}, 10(5):e0126894, 2015.

\bibitem{zhao2014fluxflow}
Jian Zhao, Nan Cao, Zhen Wen, Yale Song, Yu-Ru Lin, and Christopher Collins.
\newblock \# fluxflow: Visual analysis of anomalous information spreading on
  social media.
\newblock {\em IEEE transactions on visualization and computer graphics},
  20(12):1773--1782, 2014.

\bibitem{li2025effect}
Ming Li, Liang’an Huo, Yafang Dong, Xiaoxiao Xie, and Yingying Cheng.
\newblock Effect of individual heterogeneity on the coupled spread of
  information and disease in higher-order multiplex networks.
\newblock {\em Nonlinear Dynamics}, 113(9):10659--10679, 2025.

\bibitem{soriano2018spreading}
David Soriano-Pa{\~n}os, L~Lotero, Alex Arenas, and Jes{\'u}s
  G{\'o}mez-Garde{\~n}es.
\newblock Spreading processes in multiplex metapopulations containing different
  mobility networks.
\newblock {\em Physical Review X}, 8(3):031039, 2018.

\bibitem{keeling2010individual}
Matt~J Keeling, Leon Danon, Matthew~C Vernon, and Thomas~A House.
\newblock Individual identity and movement networks for disease
  metapopulations.
\newblock {\em Proceedings of the national academy of sciences},
  107(19):8866--8870, 2010.

\bibitem{feng2020epidemic}
Liang Feng, Qianchuan Zhao, and Cangqi Zhou.
\newblock Epidemic in networked population with recurrent mobility pattern.
\newblock {\em Chaos, Solitons \& Fractals}, 139:110016, 2020.

\bibitem{gao2022epidemic}
Shupeng Gao, Xiangfeng Dai, Lin Wang, Nicola Perra, and Zhen Wang.
\newblock Epidemic spreading in metapopulation networks coupled with awareness
  propagation.
\newblock {\em IEEE Transactions on Cybernetics}, 53(12):7686--7698, 2022.

\bibitem{shao2022epidemic}
Qi~Shao and Dun Han.
\newblock Epidemic spreading in metapopulation networks with heterogeneous
  mobility rates.
\newblock {\em Applied Mathematics and Computation}, 412:126559, 2022.

\bibitem{kar2019stability}
TK~Kar, Swapan~Kumar Nandi, Soovoojeet Jana, and Manotosh Mandal.
\newblock Stability and bifurcation analysis of an epidemic model with the
  effect of media.
\newblock {\em Chaos, Solitons \& Fractals}, 120:188--199, 2019.

\bibitem{zhu2023epidemic}
Yujie Zhu, Cong Li, and Xiang Li.
\newblock Epidemic spreading on coupling network with higher-order information
  layer.
\newblock {\em New Journal of Physics}, 25(11):113043, 2023.

\bibitem{ruan2012epidemic}
Zhongyuan Ruan, Ming Tang, and Zonghua Liu.
\newblock Epidemic spreading with information-driven vaccination.
\newblock {\em Physical Review E—Statistical, Nonlinear, and Soft Matter
  Physics}, 86(3):036117, 2012.

\bibitem{benson2016higher}
Austin~R Benson, David~F Gleich, and Jure Leskovec.
\newblock Higher-order organization of complex networks.
\newblock {\em Science}, 353(6295):163--166, 2016.

\bibitem{nie2023voluntary}
Yanyi Nie, Sheng Su, Tao Lin, Yanbing Liu, and Wei Wang.
\newblock Voluntary vaccination on hypergraph.
\newblock {\em Communications in Nonlinear Science and Numerical Simulation},
  127:107594, 2023.

\bibitem{ma2024impact}
Jinlong Ma and Peng Wang.
\newblock Impact of community networks with higher-order interaction on
  epidemic dynamics.
\newblock {\em Chaos, Solitons \& Fractals}, 180:114471, 2024.

\bibitem{wan2022multilayer}
Jinming Wan, Genki Ichinose, Michael Small, Hiroki Sayama, Yamir Moreno, and
  Changqing Cheng.
\newblock Multilayer networks with higher-order interaction reveal the impact
  of collective behavior on epidemic dynamics.
\newblock {\em Chaos, Solitons \& Fractals}, 164:112735, 2022.

\bibitem{battiston2021physics}
Federico Battiston, Enrico Amico, Alain Barrat, Ginestra Bianconi, Guilherme
  Ferraz~de Arruda, Benedetta Franceschiello, Iacopo Iacopini, Sonia K{\'e}fi,
  Vito Latora, Yamir Moreno, et~al.
\newblock The physics of higher-order interactions in complex systems.
\newblock {\em Nature Physics}, 17(10):1093--1098, 2021.

\bibitem{iacopini2019simplicial}
Iacopo Iacopini, Giovanni Petri, Alain Barrat, and Vito Latora.
\newblock Simplicial models of social contagion.
\newblock {\em Nature communications}, 10(1):2485, 2019.

\bibitem{barabasi1999emergence}
Albert-L{\'a}szl{\'o} Barab{\'a}si and R{\'e}ka Albert.
\newblock Emergence of scaling in random networks.
\newblock {\em science}, 286(5439):509--512, 1999.

\bibitem{guo2020future}
Bin Guo, Yasan Ding, Lina Yao, Yunji Liang, and Zhiwen Yu.
\newblock The future of false information detection on social media: New
  perspectives and trends.
\newblock {\em ACM Computing Surveys (CSUR)}, 53(4):1--36, 2020.

\bibitem{zhan2025measuring}
Xiu-Xiu Zhan, Chuang Liu, Zhipeng Wang, Huijuan Wang, Petter Holme, and Zi-Ke
  Zhang.
\newblock Measuring and utilizing temporal network dissimilarity.
\newblock {\em Communications Physics}, 8(1):40, 2025.

\end{thebibliography}

\end{document}